\documentstyle [11pt]{article}
\begin{document}
\begin{flushright}
PREPRINT \\
Revised\\
IC/99/51\\
US-FT/13-99\\
hep-ph/9905310

Vesrion of
December 1999
\end{flushright}

\begin{center}

{\Large \bf
Higher twists and $\alpha_s(M_Z)$ extractions
from the NNLO QCD analysis of the CCFR data for the $xF_3$
structure function.}
\vspace{0.1cm}

{\bf A.L. Kataev}

\vspace{0.1cm}

{\baselineskip=14 pt 
Institute for Nuclear Research of the Academy of Sciences
of Russia,\\ 117312 Moscow, Russia}

\vspace{0.1cm}

{\bf G. Parente}

\vspace{0.1cm}

{\baselineskip=14pt Department of Particle Physics, University
of Santiago de Compostela,\\ 15706 Santiago de Compostela, Spain}

\vspace{0.1cm}

{\bf  A.V. Sidorov}
\vspace{0.1cm}

{\baselineskip=14pt Bogoliubov Laboratory of Theoretical Physics,\\ Joint
Institute for Nuclear Research,\\ 141980 Dubna, Russia}
\end{center}
\begin{abstract}
{A  detailed  next-to-next-to-leading order (NNLO)
QCD
analysis is performed for the experimental data of the CCFR
collaboration for the $xF_3$ structure function.
Theoretical ambiguities of the results of our NNLO fits are estimated
by application of the Pad\'e resummation technique and variation
of the factorization and renormalization scales.
The
NNLO and   N$^3$LO $\alpha_s(Q^2)$ $\overline{MS}$-matching
conditions are used.
In the process of the fits we are taking account of
twist-4 $1/Q^2$-terms.
We found that the amplitude of the $x$-shape
of the twist-4 factor is  decreasing in NLO
and NNLO, though some remaining twist-4 structure seems to
retain  in  NNLO in the case when only statistical
uncertainties are taken into account.
The question of the stability of these results
to the application of the [0/2] Pad\'e resummation technique
is considered.
Our  NNLO results for $\alpha_s(M_Z)$ values,
extracted from the CCFR $xF_3$ data, are
$\alpha_s(M_Z)=
0.118 \pm 0.002 (stat)\pm 0.005 (syst) \pm 0.003(theory)$
provided the twist-4
contributions are fixed through the infrared renormalon model and
$\alpha_s(M_Z)=0.121^{+ 0.007}_{-0.010}(stat)\pm 0.005 (syst) \pm
0.003(theory)$
provided the twist-4 terms are considered as  free parameters.
}\\[5mm]

\end{abstract}

\newpage
{\bf 1. Introduction.}~

Deep-inelastic lepton-nucleon scattering (DIS) belongs to
classical and constantly studied processes in  modern
particle physics. The traditionally measurable characteristics
of $\nu N$ DIS are  SFs (structure functions) $F_2$ and $xF_3$.
It should be stressed that the
program of getting  information about the behavior of
SFs of $\nu N$ DIS is among the aims of the experimental
program of Fermilab Tevatron and CCFR/NuTeV collaboration,
in particular. The CCFRR collaboration
started to study the $\nu N$ scattering process in 1980
\cite{CCFRR}. The data for
SFs of $\nu N$ DIS, obtained by a
follower of the CCFRR collaboration,
namely, the CCFR group, were distributed among the potential users in the
beginning of 1997 \cite{Seligman}, while the final results of the original
CCFR DGLAP \cite{DGLAP} NLO analysis of these data were presented
in the journal publication of Ref.\cite{Seligman}.

This experimental information was already used in the process
of different NLO analyses, performed by CTEQ,
MRST, and GRV  groups (see Refs.\cite{CITEQ,MRST,GRV}, respectively).
Subsequent steps of performing NLO and the first NNLO analysis
of  the CCFR data with the help
of the Jacobi polynomial - Mellin moments version of the DGLAP
method
were made in Refs.\cite{KS}-\cite{KPS}
(definite stages in the development of this formalism
are described in Refs.\cite{PS}-\cite{Kretal}).

In the process of the analysis of Refs.\cite{KKPS1}-\cite{KPS},
the authors used  important
information about the NNLO corrections to the coefficient
functions \cite{WZ} and  results of the complicated
analytic calculations of
the NNLO corrections to the anomalous dimensions
of the NS moments with $n=2,4,6,8,10$ \cite{LRV1}, supplemented
with the
estimated values of the NNLO coefficients of the
anomalous dimensions of $n=3,5,7,9$ NS moments,
which were obtained in Ref.\cite{KKPS1} with the
help of
the smooth interpolation procedure, previously
proposed in Ref.\cite{PKK}. Moreover, the attempts to obtain
the shape of the twist-4 contributions and to study the predictive
abilities of the IR-renormalon (IRR) model of Ref.\cite{DW} were made
(for certain details of modeling the effects of the power-suppressed
contributions  to $xF_3$ and other
 measurable physical quantities in  the IRR
language see  Ref.\cite{MSSM} and Ref.\cite{IRR} respectively,
for more details, see  the review of Ref.\cite{Beneke}).

However,  an important question of  estimating
theoretical uncertainties of the NNLO analysis of the CCFR data
of Ref.\cite{KKPS2} was not analyzed in detail.
These uncertainties can be specified in part after taking into account

1)  differences in the definitions of  $\alpha_s(Q^2)$
matching conditions (see e.g. Refs.\cite{BW, Marciano, DV,BN}), which are
responsible
for penetrating into the energy region, characteristic of $f=5$ numbers
of flavours, where the pole of the $Z^0$-boson  manifests itself;

2)
NNLO QCD contribution
to the matching condition of  Ref.\cite{BW} (corrected in Ref.\cite{LRVS}),
the 4-loop coefficient
of the QCD $\beta$-function \cite{RVL}, which is entering
into the N$^3$LO approximation of the renormalization group evolution
equation for the Mellin moments, the N$^3$LO expression
for the coupling constant and to
the calculated N$^3$LO-term \cite{CKS}
of the matching condition;

3) theoretical
uncertainties due to  non-calculated N$^3$LO contributions to
the coefficient functions and the anomalous-dimension functions;

4) in the NNLO it is also desirable to carefully analyse  the
dependence  of the results obtained on the choice of the Jacobi
polynomial parameters $\alpha$ and $\beta$ (see definitions below).
These parameters are entering into the theoretical expression for the
reconstructed structure function (in the NLO the problem of their
fixation
was studied in detail in Ref.\cite{Kretal}). This analysis
is of relevant importance in view of
doubts  in the applicability of the Jacobi evolution method
(see Ref.\cite{GR}), which, however, were immediately dispelled
in Ref.\cite{Shaw};

5) last, but not least, uncertainties are related to
problems of the sensitivity of the obtained results to
the choice of  factorizatation and renormalization scales.
It is worth  noting that these problems are in
relation with fixing  the ambiguities due to
already calculated and uncalculated N$^3$LO QCD effects, which can
be modelled using
the Pad\'e resummation technique.

This work is devoted to the  analysis of  important problems
outlined above and to a more detailed determination of  values
of $\alpha_s(M_Z)$ and the $x$-shape of the
twist-4 power-suppressed term in available orders of perturbative QCD
taking into account the effects
listed above. We  supplement the NNLO fits of
Ref.\cite{KKPS2} by the N$^3$LO analysis based on the
application of the Pad\'e resummation technique (for a review, see
Ref.\cite{Pade}), developed in QCD in a
definite form in Refs.\cite{SEK,EGKS}, and considered previously as
a possible method of fixing theoretical uncertainties in the analysis of
DIS data in Ref.\cite{Sid}. It should be stressed that
{\it a posteriori} this technique gives  results similar  to
those obtained by
different methods of fixing scale-scheme
dependence ambiguities (compare the results of Ref.\cite{KatSt} with
the results
of Refs.\cite{SEK,EGKS} obtained by using  the Pad\'e resummation technique).
Thus, our analysis could be considered as an attempt to estimate
perturbative QCD uncertainties beyond the NNLO level. Moreover, it could give
us a hint whether the outcomes of our NNLO fits, related to perturbative
and non-perturbative sectors, are affected  upon  including  the
explicitly calculated and estimated N$^3$LO QCD effects.

Another new important  ingredient of our analysis, discussed in brief
in Ref.\cite{KPS99}, is the analysis of the problem of the initial
scale choice. In particular, we will demonstrate that due to
unnaturally large NNLO corrections to the renormalization-group
improved $n$=2 Mellin moment
of   $xF_3$ ,  it is
essential to choose the value of the initial point in a vicinity
of the scale
$Q_0^2$=20 $GeV^2$.

{\bf 2. The theoretical background of the QCD analysis.}~

Let us  define the Mellin moments
for  NS SF $xF_3(x,Q^2)$:
\begin{equation}
M_n^{NS}(Q^2)=\int_0^1 x^{n-1}F_3(x,Q^2)dx
\end{equation}
where $n=2,3,4,...$. These moments
obey the following renormalization group equation
\begin{equation}
\bigg(\mu\frac{\partial}{\partial\mu}+\beta(A_s)\frac{\partial}
{\partial A_s}
+\gamma_{NS}^{(n)}(A_s)\bigg)
M_n^{NS}(Q^2/\mu^2,A_s(\mu^2))=0
\label{rg}
\end{equation}
where $A_s=\alpha_s/(4\pi)$.
The renormalization group functions are defined as
\begin{eqnarray}
\mu\frac{\partial A_s}{\partial\mu}=\beta(A_s)=-2\sum_{i\geq 0}
\beta_i A_s^{i+2}~~\nonumber \\
\mu\frac{\partial ln
Z_n^{NS}}{\partial\mu}=\gamma_{NS}^{(n)}(A_s) =\sum_{i\geq 0}
\gamma_{NS}^{(i)}(n) A_s^{i+1}
\end{eqnarray}
where
$ Z_n^{NS} $
are renormalization constants of the corresponding
NS operators.  The solution of the renormalization group
equation
can be represented in the following form :
\begin{equation}
\frac{ M_{n}^{NS}(Q^2)}{M_{n}^{NS}(Q_0^2)}=
exp\bigg[-\int_{A_s(Q_0^2)}^{A_s(Q^2)}
\frac
{\gamma_{NS}^{(n)}(x)}{\beta(x)}dx\bigg]
\frac{C_{NS}^{(n)}(A_s(Q^2))}
{C_{NS}^{(n)}(A_s(Q_0^2))}
\label{mom}
\end{equation}
where $M_n^{NS}(Q_0^2)$ is a phenomenological quantity related to the
initial- scale dependent factor.
At a fixed momentum transfer $Q_0^2$, it will be parameterized
in the simple form
\begin{equation}
M_n^{NS}(Q_0^2)=\int_0^1 x^{n-2} A(Q_0^2)x^{b(Q_0^2)}(1-x)^{c(Q_0^2)}
(1+\gamma(Q_0^2)x)dx
\end{equation}
with $\gamma \neq 0$ or $\gamma=0$. It is
identical to the form used by the CCFR collaboration \cite{Seligman}.
In principle, following the
models of parton distributions used in Refs.\cite{MRST,GRV},
one can  in Eq.(5) add a term proportional to
$\sqrt{x}$. However, since this term is  important only in the region of
rather
small $x$, we  neglect it in   our analysis.

In N$^3$LO,
the coefficient function $C_{NS}^{(n)}$  can be defined as
\begin{equation}
C_{NS}^{(n)}(A_s)=1+C^{(1)}(n)A_s+C^{(2)}(n)A_s^2+C^{(3)}(n)A_s^3,
\end {equation}
while the corresponding expansion of the anomalous-dimension term
is
\begin{equation}
exp\bigg[-\int^{A_s(Q^2)}
\frac{\gamma_{NS}^{(n)}(x)}{\beta(x)}dx\bigg]=
\big(A_s(Q^2)\big)^{\gamma_{NS}^{(0)}(n)/2\beta_0}
\times AD(n,A_s)
\end{equation}
where
\begin{equation}
AD(n,A_s)=
[1+p(n)A_s(Q^2)
+q(n)A_s(Q^2)^2+r(n)A_s(Q^2)^3]
\label{an}
\end{equation}
and  $p(n)$, $q(n)$ and $r(n)$ have the following form:
\begin{equation}
p(n)=\frac{1}{2}\bigg(\frac{\gamma_{NS}^{(1)}(n)}{\beta_1}-
\frac{\gamma_{NS}^{(0)}(n)}{\beta_0}\bigg)\frac{\beta_1}{\beta_0}
\end{equation}
\begin{equation}
q(n)=\frac{1}{4}\bigg( 2p(n)^2+
\frac{\gamma_{NS}^{(2)}(n)}{\beta_0}+\gamma_{NS}^{(0)}(n)
\frac{(\beta_1^2-\beta_2\beta_0)}{\beta_0^3}-\gamma_{NS}^{(1)}(n)
\frac{\beta_1}{\beta_0^2}\bigg)
\end{equation}
\begin{eqnarray}
r(n)=\frac{1}{6}\bigg(-2p(n)^3+6p(n)q(n)+
\frac{\gamma_{NS}^{(3)}(n)}{\beta_0}-\frac{\beta_1\gamma_{NS}^{(2)}(n)}
{\beta_0^2} \\ \nonumber
-\frac{\beta_2\gamma_{NS}^{(1)}(n)}{\beta_0^2}+
\frac{\beta_1^2\gamma_{NS}^{(1)}(n)}{\beta_0^3}
-\frac{\beta_1^3\gamma_{NS}^{(0)}(n)}{\beta_0^4}
-\frac{\beta_3\gamma_{NS}^{(0)}(n)}{\beta_0^2}+
\frac{2\beta_1\beta_2\gamma_{NS}^{(0)}(n)}{\beta_0^3}\bigg)
\end{eqnarray}
The coupling constant $A_s(Q^2)$ can be expressed
in terms
of the inverse powers of
$L=\ln(Q^2/\Lambda_{\overline{MS}}^2)$
as $A_s^{NLO}=A_s^{LO}+\Delta A_s^{NLO}$,
$A_s^{NNLO}=A_s^{NLO}+\Delta A_s^{NNLO}$ and
$A_s^{N^3LO}=A_s^{NNLO}+\Delta A_s^{N^3LO}$, where
\begin{eqnarray}
A_s^{LO}&=&\frac{1}{\beta_0
L} \\ \nonumber
\Delta A_s^{NLO}&=&
-\frac{\beta_1 ln(L)}{\beta_0^3 L^2}
\end{eqnarray}
\begin{equation}
\Delta A_s^{NNLO}=\frac{1}{\beta_0^5 L^3}[\beta_1^2 ln^2 (L)
-\beta_1^2 ln(L) +\beta_2\beta_0-\beta_1^2]
\end{equation}
\begin{eqnarray}
\Delta A_s^{N^3LO}=\frac{1}{\beta_0^7 L^4}[\beta_1^3 (-ln^3 (L)
+\frac{5}{2}ln^2 (L)
+2ln(L)-\frac{1}{2})
\\ \nonumber
-3\beta_0\beta_1\beta_2 ln(L)
+\beta_0^2\frac{\beta_3}{2}]~.
\end{eqnarray}

Notice that in our normalization the numerical expressions for
$\beta_0$, $\beta_1$, $\beta_2$ and $\beta_3$ read
\begin{eqnarray}
\beta_0&=&11-0.6667f \nonumber \\
\beta_1&=&102-12.6667f \nonumber \\
\beta_2&=&1428.50-279.611f+6.01852f^2 \nonumber \\
\beta_3&=&29243.0-6946.30f+405.089f^2+1.49931f^3
\end{eqnarray}
where the expression for $\beta_3$ was obtained in
Ref.\cite{RVL}. The inverse-log expansion for
$\Delta A_s^{N^3LO}$ which incorporates the information
about the coefficient $\beta_3$ was presented in Ref.\cite{CKS}.

A few words are to be said about the  approximation used for the
anomalous-dimension function $\gamma_{NS}^{(n)}(A_s)$.
The analytic expression for its one-loop coefficient
is well-known: $\gamma_{NS}^{(0)}(n)=
(8/3)[4\sum_{j=1}^{n}(1/j)-2/n(n+1)-3]$.
In the cases of both $F_2$ and $xF_3$,  numerical expressions
for $\gamma_{NS}^{(1)}(n)$-coefficients are given in Table 1.

\begin{center}
\begin{tabular}{||r|c|c|c|c|c|}
\hline
n  & $\gamma_{NS,F_2}^{(1)}(n)$&$\gamma_{NS,F_3}^{(1)}(n)$
& $\gamma_{NS}^{(2)}(n)$& $\gamma_{NS}^{(3)}(n)|_{[1/1]}$
& $\gamma_{NS}^{(3)}(n)|_{[0/2]}$ \\
\hline
          2&   71.374&      71.241 &           612.1 & 5259 & 5114 \\
          3&  100.801&      100.782 &          837.4 & 6959 & 6900  \\
          4&  120.145&      120.140 &         1005.8 & 8421 & 8414  \\
          5&  134.905&      134.903 &         1135.8 & 9563 & 9562  \\
          6&  147.003&      147.002 &         1242.0 & 10493 & 10482 \\
          7&  157.332&      157.332 &         1334.0 & 11310 & 11280  \\
          8&  166.386&      166.386 &         1417.5 & 12077 & 12012 \\
          9&  174.468&      174.468 &         1493.5 & 12784 & 12706  \\
         10&  181.781&      181.781 &         1559.0 & 13370 & 13271 \\
         11&  188.466&      188.466 &          ?  & ? & ? \\
         12&  194.629&      194.629 &         ?   & ? & ? \\
         13&  200.350&      200.350 &         ?   & ? & ? \\
         14&  205.689&      205.689 &         ?   & ? & ? \\
\hline
\end{tabular}

\end{center}
{{\bf Table 1.} The used numerical expressions for  NLO
and NNLO coefficients of  anomalous dimensions of  moments of
the NS SFs at $f=4$ number of flavours and the N$^3$LO Pad\'e estimates.}
\vspace{0.5cm}

These results are normalized to the case of
$f=4$ numbers of active flavours. In the same Table
we present
the numerical
expressions for $\gamma_{NS}^{(2)}(n)$, used in the process of  fits.
In the cases of $n=2,4,6,8,10$ they follow from the
explicit calculations of $\gamma_{NS,F_2}^{(2)}(n)$-terms \cite{LRV1},
normalized to $f=4$, while the $n=3,5,7,9$ numbers are fixed by using
the smooth interpolation procedure originally proposed in Ref.\cite{PKK}.
Note in advance, that since $\gamma_{NS,F_3}^{(2)}(n)$-coefficients
differ from $\gamma_{NS,F_2}^{(2)}(n)$-terms, though by  presumably small
additional contributions (for discussions, see Ref.\cite{KKPS1}),
it would be interesting to verify the accuracy of the  expression
for $\gamma_{NS}^{(2)}(n)$, used in the process of our  NNLO
$xF_3$ fits, by  explicit analytic calculations of the
NNLO contributions to  anomalous dimensions of  odd moments
of the $xF_3$ structure function.

Let us now describe the procedure of fixing other theoretical
uncertainties.
After the work of Ref.\cite{SEK} it becomes rather popular to
model the effects of higher order terms of perturbative
series in QCD by the expanded Pad\'e approximants.

In the framework of this
technique, the values of  terms $C^{(3)}(n)$  and $r(n)$  could be
expressed as

\begin{eqnarray}
Pade~[1/1]~: && C^{(3)}(n)=[C^{(2)}(n)]^2/C^{(1)}(n)        \\
&&   r(n)=q(n)^2/p(n) \\
Pade~[0/2]~: &&
C^{(3)}(n)=2C^{(1)}(n)C^{(2)}(n)     -[C^{(1)}(n)]^3
\\
&&   r(n)= 2p(n)q(n)-[p(n)]^3
\end{eqnarray}

The numerical values for $p(n)$ and $q(n)$, obtained from the
results of Table 1 and definitions of Eqs.(9)-(11), together
with the values of the coefficients $C^{(1)}(n)$ and $C^{(2)}(n)$
(which  come from the calculations of Ref.\cite{WZ}), are presented
in Table 2.

\begin{center}
\begin{tabular}{||r|c|c|c|c|c|c|c|c||} \hline
n&
$p(n)$ & $q(n)$ & $r(n)|_{[1/1]}$ & $r(n)|_{[0/2]}$ &
$C^{(1)}(n)$ & $C^{(2)}(n)$ & $C^{(3)}(n)|_{[1/1]}$ &
$C^{(3)}(n)|_{[0/2]}$ \\ \hline
2& 1.646 & 4.232  & 10.829 &  9.476 & -1.778 & -47.472 & -1268 &
174
     \\
3& 1.941 & 4.774  & 11.738    & 11.218 & 1.667 & -12.715 & 97 & -47
          \\
4& 2.050  & 5.546      & 15.003  & 14.123 & 4.867 & 37.117 & 283& 246
 \\
5&  2.115 &   6.134      & 17.790 & 16.486 & 7.748 & 95.408 & 1175 &
1013                    \\
6 & 2.165 &  6.595      & 20.087 & 18.407 & 10.351 & 158.291 & 2421 &
2168
          \\
7&  2.210 &   7.039     & 22.421 & 20.318 & 12.722 & 223.898 &
3940 & 3638                     \\
8&  2.252 &  7.525 & 25.138 & 22.471 & 14.900 & 290.884 & 5679 &
5360
                    \\
9& 2.294 &  8.018     &  28.027 &  24.715 & 16.915 & 358.587 &
7602 & 7291
         \\
10 & 2.334 &  8.375     &  30.049 & 26.382 & 18.791 & 426.442 & 9677
&         9391                     \\
\hline
\end{tabular}
\end{center}

{{\bf Table 2.} The  values for  NLO and NNLO
QCD contributions used in our fits and the N$^3$LO Pad\'e estimates.}
\vspace{0.5cm}

In the same Table, we also give the estimates for $r(n)$
and $C^{(3)}(n)$, obtained by using the expanded [1/1] and [0/2]
Pad\'e approximants formulae of Eqs.(16)-(19).
For  completeness,
in the last two columns of Table 1 we also present the estimates for
N$^3$LO contributions to the anomalous dimension function
$\gamma_{NS}^{(n)}(A_s)$, obtained with the help of the expanded [1/1] and
[0/2] Pad\'e approximants.
One can see that the results of applications of [1/1] and [0/2]
Pad\'e approximants for $\gamma_{NS}^{(3)}(n)$ are almost identical
to each other.

Using the numbers presented in Table 1, one can construct Pad\'e
motivated expressions for $r(n)$  by substituting the estimates
for $\gamma_{NS}^{(3)}(n)$ into  Eq.(11). It should be stressed, that the
obtained  estimates for $r(n)$  qualitatively agree with
the ones presented in Table 2 within the ``Pad\'e world'' only, namely
only in the case of application, in Eq.(11), of the [1/1] or [0/2] Pad\'e
estimate for the four-loop coefficient of the QCD $\beta$-function $\beta_3$.
However, in the case of $f=4$ the direct application of the [1/1] and [0/2]
Pad\'e approximants underestimates the
calculated value of $\beta_3$ by  a factor of over
2.5 $(\beta_3|_{[1/1]}\approx 3217;\beta_3|_{[0/2]}\approx3058).$
In view of this, the application of Eq.(11) with the Pad\'e estimated values
of $\gamma_{NS}^{(3)}(n)$ and the explicit expression for $\beta_3$-coefficient
give estimates of $r(n)$, drastically different from the ones
presented in Table 2 (for example, for the case of application
of [0/2] P\'ade estimates it gives $r(2)\approx 16.6$,....,$r(10)\approx
49.3$).

It is already known that the accuracy of  estimates of the
N$^3$LO coefficient of the QCD $\beta$-function can be improved
by some additional fits of the polynomial dependence of $\beta_3$
on the number
of flavours $f$ and by applying the asymptotic Pad\'e approximant
(APAP) formula
\cite{EKSbet}. Therefore, it might be interesting to consider
the possibility of making
Pad\'e estimates of N$^3$LO contributions
to $\gamma_{NS}(A_s)$ (see Table 1) more theoretically motivated.
Analogous steps were already done
in Ref.\cite{Padad}
in the analysis of the status of N$^3$LO Pad\'e
estimates for the anomalous dimension
function of quark mass.
The agreement of the obtained estimates with the   calculated
four-loop QCD  results of Ref.\cite{gamam,gam2}  turned
out to be reasonable. One can hope that the application
of  similar procedure for the APAP estimates of
$\gamma_{NS}^{(3)}$-terms and the substitution of the results
obtained in Eq.(11) (together with the explicit expression
for the $\beta_3$-term) might improve the agreement with the estimates
presented in Table 2. At this
step we consider the estimates presented in Table 2 as  suitable
results for modelling
the unknown effects of the N$^3$LO corrctions, which
depend on  the N$^3$LO expression
for the coupling constant $A_s$.

Within this approach, the uncertainties of the results of
NNLO fits can be estimated by modelling $q(n)$ and $C^{(2)}(n)$ by
using the [0/1] Pad\'e approximants, which give $q(n)|_{[0/1]}=[p(n)]^2$ and
$C^{(2)}(n)|_{[0/1]}=[C^{(1)}(n)]^2$. The estimated values of
$q(n)|_{[0/1]}$ are correct in sign for $n\geq2$, while for $C^{(2)}(n)$
the same feature takes place in the case of $n\geq 4$ moments.
Moreover, for $n\geq 4$ the relative values of  ratios
$q(n)|_{[0/1]}/q(n)$ are varying from 1.3 to 1.5, while
similar ratios for  NLO contributions to the
coefficient function $C^{(2)}(n)|_{[0/1]}/C^{(2)}(n)$ are changing
from 1.6 at $n=4$ to 1.2 at $n=10$. This precision seems to be
rather acceptable for
the [0/1] Pad\'e estimates, which in the case of each concrete fixed value
of  $n$ are
based on one input term
of the corresponding perturbative series.

It should also be  stressed that the uncertainties in
values of $r(n)$ are not so important, since the results of our fits
are more sensitive to the form of predictions of the
Pad\'e approximations  for the
N$^3$LO contributions to the
coefficient function (namely, $C^{(3)}(n)$-terms).

From the  results presented in Table 2   one can conclude
that the theoretical series for $C_{NS}^{(n)}$ for large $n$ $(n\geq 4)$,
relevant to the  behavior of $xF_3(x,Q^2)$  in the intermediate
and large $x$-region, probably have sign constant
structure with asymptotically increasing positive coefficients.
Therefore, the applications of the expanded [1/1] and [0/2] Pad\'e
approximants for
estimating  the terms $C^{(3)}(n)$ with $n\geq 4$
(which in both cases
have the same positive sign and the same order of magnitude)
might be considered as the useful ingredient for the N$^3$LO Pad\'e-motivated
fits.

However,
in the cases  of  coefficient functions of
$n=2,3$ moments of  $xF_3$ 
our intuition suggest nothing about
the sign and order of magnitude of the third term in perturbative
series  $C_{NS}^{(2)}(A_s)=1-1.78A_s-47.47A_s^2$ and
$C_{NS}^{(3)}(A_s)=1+1.67A_s-12.71A_s^2$.
Indeed, in these two cases the manipulations with [1/1] and [0/2]
Pad\'e approximants  give drastically different estimates
for the terms $C^{(3)}(n)$ which for $n=2,3$ differ
both in sign and size (see Table 2).
It is  possible that this feature is related to the fact
that for $n=2,3$ the coefficients of $C_{NS}^{(n)}(A_s)$  have
no immediate $(+1)^{m} m!$ growth, but exhibit
some zigzag structure manifesting itself in the cases of definite
perturbative series of quantum field theory models (for discussions, see
e.g. Ref.\cite{BL}). This might give  additional theoretical
uncertainties of modelling higher-order perturbative QCD predictions
for $xF_3(x,Q^2)$ in the region of relatively small  $x$.

In view of the questionable asymptotic behavior of the NNLO
series for coefficient functions of the NS moments with low $n$
($n=2,3$), we  also
use the  idea of Ref.\cite{Sid} and
consider 
non-expanded Pad\'e approximants in the process of
analysis of the DIS data.

Let us recall that
the corresponding non-expanded [1/1] Pad\'e approximants can be defined as
\begin{equation}
AD(n, A_s)|_{[1/1]}=\frac{1+a_1^{(n)}A_s}{1+b_1^{(n)}A_s}
\end{equation}
\begin{equation}
C_{NS}^{(n)}(A_s)|_{[1/1]}=\frac{1+c_1^{(n)}A_s}{1+d_1^{(n)}A_s}
\end{equation}
where $a_1^{(n)}=\bigg([p(n)]^2-q(n)\bigg)/p(n)$, $b_1^{(n)}=-q(n)/p(n)$
and $C_1^{(n)}=\bigg([C^{(1)}(n)]^2-C^{(2)}(n)\bigg)/C^{(1)}(n)$,
$d_1^{(n)}=-C^{(2)}(n)/C^{(1)}(n)$.

The explicit expressions
for the non-expanded [0/2] Pad\'e approximants read:
\begin{equation}
AD(n,A_s)|_{[0/2]}=\frac{1}{1+b_1^{(n)}A_s+b_2^{(n)}A_s^2}
\end{equation}
\begin{equation}
C_{NS}^{(n)}(A_s)|_{[0/2]}=\frac{1}{1+d_1^{(n)}A_s+d_2^{(n)}A_s^2}
\end{equation}
where $b_1^{(n)}=-p(n)$, $b_2^{(n)}=p(n)^2-q(n)$, $d_1^{(n)}=-C^{(1)}(n)$
and $d_2^{(n)}=[C^{(1)}(n)]^2-C^{(2)}(n)$.
Since we consider the applications of both [1/1] and [0/2] Pad\'e
approximants as  attempts to model the behavior of
the perturbative
series for the NS Mellin moments beyond the NNLO level, we  use,
in Eqs.(20)-(23),  the N$^3$LO expression for the coupling constant $A_s$,
defined  through Eqs.(12)-(14). It is worth  mentioning here that
quite recently   the expanded and non-expanded
Pad\'e approximants
were successfully used to study     the N$^3$LO approximation
of the ground state energy in quantum mechanics \cite{PP} and
the behavior of the $\beta$-function
for the quartic Higgs coupling in the Standard Electroweak Model
\cite{Durand}.

The next step is the reconstruction of the structure
function $xF_3(x,Q^2)$  with
both target mass corrections and twist-4 terms taken into account.
The reconstructed SF can be expressed as:
\begin{eqnarray}
xF_{3}^{N_{max}}(x,Q^2)&=&
w(\alpha,\beta)(x)
\sum_{n=0}^{N_{max}}
\Theta_n ^{\alpha , \beta}
(x)\sum_{j=0}^{n}c_{j}^{(n)}{(\alpha ,\beta )}
M_{j+2,xF_3}       \left ( Q^{2}\right ) \\ \nonumber
&&+\frac{h(x)}{Q^2}
\label{Jacobi}
\end{eqnarray}
where $\Theta_n^{\alpha,\beta}$ are the Jacobi polynomials, 
$c_j^{(n)}(\alpha,\beta)$  contain 
$\alpha$ and $\beta$ dependent Euler $\Gamma$-functions where 
$\alpha,\beta$ are the Jacobi polynomials parameters, fixed by
the minimization of the error in the reconstruction of
the SF, and $w(\alpha,\beta)=x^{\alpha}(1-x)^{\beta}$ is the corresponding
weight function.
To take into account the target mass corrections, the
Nachtmann moments
\begin{eqnarray}
M_{n,xF_3}\rightarrow
M_{n,xF_3}^{TMC}(Q^2)=\int_{0}^{1}\frac{dx\xi^{n+1}}{x^2}F_3(x,Q^2)
\frac{1+(n+1)V}{(n+2)},
\label{f3}
\end{eqnarray}
can be used,
where
$\xi=2x/(1+V)$, $V=\sqrt{1+4M_{nucl}^2x^2/Q^2}$ and
$M_{nucl}$ is the mass of a nucleon.
However, to simplify the analysis, it is convenient
to expand
equation  (\ref{f3})  into a series in
powers of $M_{nucl}^2/Q^2$ \cite{GEORGI}. Taking into account the
order $O(M_{nucl}^4/Q^4)$ corrections, we get

\begin{eqnarray}
M_{n,xF_3}^{TMC}(Q^2)&=&M_{n,xF_3}^{NS}(Q^2)+\frac{n(n+1)}{n+2}
\frac{M_{nucl.}^2}{Q^2}
M_{n+2,xF_3}^{NS}(Q^2) \\ \nonumber
&&+\frac{(n+2)(n+1)n}{2(n+4)}
\frac{M_{nucl.}^4}{Q^4}M_{n+4,xF_3}^{NS}(Q^2)
+O(\frac{M_{nucl}^6}{Q^6}),
\label{m3}
\end{eqnarray}

We have checked that   the order
$O(M_{nucl}^4/Q^4)$ terms in Eq.(26)
have a rather small effect in the process of
concrete fits. Therefore, in what
follows we will use only the first two terms in the r.h.s. of
Eq.(26).

The form of  twist-4 contributions $h(x)$ in Eq.(24) was
first fixed as
\begin{equation}
\frac{h(x)}{Q^2}=w(\alpha,\beta)\sum_{n=0}^{N_{max}} \Theta_n^{\alpha,\beta}(x)
\sum_{j=0}^{(n)}c_j^{(n)}(\alpha,\beta)M_{j+2,xF_3}^{IRR}(Q^2)
\end{equation}
where 
\begin{equation}
M_{n,xF_3}^{IRR}(Q^2)=\tilde{C}(n)M_{n,xF_3}^{NS}(Q^2)
\frac{A_2^{'}}{Q^2}+
O(\frac{1}{Q^4})
\end{equation}
with $A_2^{'}$ taken as a free parameter and $\tilde{C}(n)$ defined
following the IRR model estimates of Ref.\cite{DW}
as $\tilde{C}(n)=-n-4+2/(n+1)+4/(n+2)+4S_1(n)$ $(S_1(n)=\sum_{j=1}^{n}
1/j)$.
It should be stressed that    the
multiplicative
QCD expression  $M_{n,xF_3}^{NS}(Q^2)$ in Eq.(28), generally speaking
different
from the intrinsic coefficient function of the twist-4
contribution,
leads to  theoretical
uncertainties
in  the contributions of higher-order QCD
corrections to the twist-4 part of  $xF_3(x,Q^2)$.
This could provide  additional  theoretical errors
in the studies of the  status of the IRR-model predictions for the
twist-4
terms in  NNLO.

To analyze this question at a more definite theoretical level,
it is instructive to model the function
$h(x)$  by  additional free parameters of the fits, not related to
the IRR-model predictions.

We will  estimate the
uncertainties of the values of $\Lambda_{\overline{MS}}^{(4)}$,
and thus  $\alpha_s(M_Z)$,
by studying the factorization and renormalization  scale dependence of
the outcomes of the fits. We will also analyze the stability
of  extracted values of the
IRR-model parameter $A_2^{'}$ and the twist-4 function
$h(x)$  to  the  explicitly
calculated N$^3$LO QCD
corrections  and other unknown N$^3$LO terms (modelled
with the help of
the Pad\'e resummation technique) included into the fits of concrete data.
Our aim will  also be the study of the influence
of  the choice of the initial  scale
on the results of Ref.\cite{KKPS2} and especially on those,
which  describe  the $x$-shape of  $h(x)$ for  $xF_3$ 
within the method adopted by us.

{\bf 3 (a). The analysis of the experimental data:
the extraction of $\Lambda_{\overline{MS}}^{(4)}$ vs $\alpha_s$ value.}

The  results for our NLO and NNLO fits,
made for the case of  number of active flavours $f=4$, are
presented in Table 1 of Ref.\cite{KKPS2}, where  the
values of the  parameters for the model of  $xF_3$ 
$A,b,c,\gamma\neq 0$ (related to the parton distribution
parameters)
are also given. The results of Ref.\cite{KKPS2} were obtained by using the
fixed value of the initial point $Q_0^2=5~GeV^2$ and the
fixed
weight function of the Jacobi polynomials reconstruction formula
of Eq.(24),
namely $x^{0.7}(1-x)^3$. Note, that this form
is similar to the $x$-shape of the
NS structure function itself. Indeed, the  value of
the parameters $\beta=3$ is   in agreement with the
quark-counting rules of Ref.\cite{QQR}, while the value $\alpha=0.7$
is close
to  the value of the parameter, which describe the
Regge theory behaviour of  NS SF at  small $x$.

In this
section, we will present more definite arguments in
favour of the used form of the Jacobi polynomial weight
function and will study the initial $Q_0^2$- scale dependence
of the results for $\Lambda_{\overline{MS}}^{(4)}$, extracted
in different orders of perturbation theory.

We will also  construct
the N$^3$LO $Q^2$-evolution equations for the Mellin moments
using the Pad\'e approximants,  written down both in the expanded
and non-expanded forms. In the process
of these ``approximate'' N$^3$LO fits, the
explicit N$^3$LO expression for
the QCD running coupling constant $A_s$, defined in Eqs.(12)-(15), will
be used.
Thus, from the fits  we obtain  the N$^3$LO estimates
of the parameter
$\Lambda_{\overline{MS}}^{(4)}$ (and therefore $\alpha_s(M_Z)$), and of the
common factor $A_2^{'}$ of the IRR model.
The comments on  attempts to apply the scheme-invariant analysis
for  estimating  the renormalization-scheme dependence of the
results obtained will  also be presented.

It should be stressed that despite the general theoretical preference
of applications of the diagonal Pad\'e approximants
(for a recent analysis, see, e.g., Ref.\cite{diagonal}),
the N$^3$LO [1/1] Pad\'e
approximant description of the CCFR'97 experimental data turned out
to be not acceptable in our case, since  it produces a rather high value
of $\chi^2$: $\chi^2/{nep}>2$ (where $nep=86$ is the number
of  experimental points, taken into account in the case of the cut
$Q^2>5~GeV^2$).
However, the application of [0/2] Pad\'e approximants produced
reasonable results.
We think that the non-applicability of the [1/1]
Pad\'e method in the process of fitting CCFR $xF_3$ data with the help of the
Jacobi polynomial approach can be related to  a rather
large value of the ratio $[C^{(3)}(2)/C^{(2)}(2)]|_{[1/1]}$
in the expression
for the NS moment $M_{2,xF_3}^{NS}$ (see Table 2).

A similar
effect of the preference of the [0/2] Pad\'e approximant analysis
over the [1/1] one was found in Ref.\cite{EGKS} from
the comparison
of  QCD
theoretical predictions for the polarized Bjorken sum rule (which
are closely related to the QCD predictions for the first
moment of  $xF_3$, namely,
for the Gross-Llewellyn Smith sum rule)
with the available experimental data.

Considering the problem of  minimization of the dependence of the
results
of the fits  on  free
parameters $\alpha$, $\beta$,
we found several minima on the ($\alpha$, $\beta$)- plane at
$Q_0^2=5~GeV^2$:
\begin{enumerate}
\item
Minimum A: $\alpha/\beta \approx$-0.6/0.55.

At this minimum we got  reasonable LO, NLO and NNLO
values of $\Lambda_{\overline{MS}}^{(4)}$ for $N_{max}=6$.
However, the appearance of this minimum strongly
depends on the number of moments taken into account.
For example, in the case of $N_{max}=10$
we were unable to find this minimum at  LO and NLO
, so we  consider this minimum as a spurious one;
\item
Minimum B : $\alpha/\beta\approx$-0.5/-0.9.

This minimum  appears in LO and NLO.
However, this minimum does not appear in NNLO,
so we consider it as  non-applicable for our
NNLO fits.
\item
Minimum C: $\alpha/\beta\approx$0.8/1.3.

For $N_{max}=6$ the LO and NLO values of $\Lambda_{\overline{MS}}^{(4)}$
can be obtained. However, in NNLO this minimum
does not manifest itself. Moreover, it  disappears in LO and NLO
for the case of $N_{max}=10$. Therefore, we  consider it as the
spurious one also.
\item
Minimum D: $\alpha/\beta\approx$ 0.6/-0.99.

It should be stressed that in LO and NLO this minimum  appears
only for $N_{max}=6$ and  disappears for $N_{max}=10$.
Moreover, the obtained value of $\beta$ results in the
unnatural singular
$1/(1-x)$ behaviour of the Jacobi polynomial weight function
$w(\alpha,\beta)$. In view of this we consider this minimum as the
unphysical one.
\item
Minimum E: $\alpha/\beta\sim$ 0.7/3.0.

It is the minimum at which we  worked earlier in
Refs.\cite{KKPS2,KPS}.
At  this minimum the results of   LO and NLO  fits
are in agreement with the ones obtained in Refs.\cite{KKPS2,KPS}.
It should be noted that in LO and NLO this
minimum is stable due to variation of $Q_0^2$  and to the inclusion of
higher Mellin moments into the reconstruction formulae of Eq.(24) and
Eq.(27) (we  checked this statement,
repeating the fits for $N_{max}=10$).
In NNLO this minimum  appears at $Q_0^2$ higher than $10~GeV^2$.
\end{enumerate}

Since the values of  parameters $\alpha$, $\beta$
at  Minimum E are  identical  to the
initially considered ones( $\alpha=0.7$, $\beta=3$) and in view of the
stability
of the results of the LO and NLO fits to the value of the initial
scale $Q_0^2$, we consider this minimum  as the physical one and will
work in its vicinity , fixing $\alpha=0.7$ and $\beta=3$
as earlier.

To study the  dependence from the choice of the initial
scale
in more detail and thus to
check the reliability
of the results, obtained in Refs.\cite{KKPS2,KPS},
we performed  LO, NLO and NNLO fits
for different values of the initial scale
without taking
account of twist-4 contributions, but with target mass corrections
included (see Ref.\cite{KPS99}).
The results for $\Lambda_{\overline{MS}}^{(4)}$ are presented in Table 3.

\begin{center}
\begin{tabular}{||c|c|c|c|c|c|c|}
\hline
$Q_0^2$ ($GeV^2$) & 5 & 8 & 10 & 20 & 50 & 100 \\
\hline
LO  & 266$\pm$35 & 266$\pm$35 & 265$\pm$34 & 264$\pm$35 & 264$\pm$36 &
263$\pm$36 \\

LO$^*$ & 382$\pm$38 & 380$\pm$41 & 380$\pm$40 & 379$\pm$46 & 378$\pm$43 &
377$\pm$42 \\

NLO & 341$\pm$30 & 340$\pm$40 & 340$\pm$35 & 339$\pm$36 & 337$\pm$34 &
337$\pm$37 \\

NLO$^*$ & 322$\pm$29 & 321$\pm$33 & 321$\pm$33 & 320$\pm$34 & 319$\pm$36 &
318$\pm$36 \\

NNLO & 293$\pm$30 & 312$\pm$33 & 318$\pm$33 & 326$\pm$35 & 326$\pm$36 &
325$\pm$36\\
\hline

NNLO$^*$ & 284$\pm$28 & 312$\pm$33 & 318$\pm$33 & 326$\pm$35 & 326$\pm$36 &
325$\pm$36 \\

N$^3$LO [0/2] & 293$\pm$29 & 323$\pm$32 & 330$\pm$35 & 335$\pm$37 & 326$\pm$
36 & 319$\pm$35 \\
\hline
\end{tabular}
\end{center}

{{\bf Table 3.} The $Q_0^2$ dependence of $\Lambda_{\overline{MS}}^{(4)}$
[MeV]. The LO$^*$ means that in the LO-fits NLO $\alpha_s$ is used;
NLO$^*$ (NNLO$^*$) indicates that in the NLO (NNLO) fits
NNLO (N$^3$LO) $\alpha_s$ is used. The N$^3$LO [0/2] marks the
results of the expanded [0/2] Pad\'e fits with $\alpha_s$ defined
in N$^3$LO.}

\hspace{0.5cm}

One can see that the LO and NLO results are stable to the variation of
$Q_0^2$. The results of LO$^*$ fits are higher than the LO ones, and
from this level   other perturbative QCD effects
tend to decrease the values of $\Lambda_{\overline{MS}}^{(4)}$ up
to the level of the NNLO$^*$-fits.

The NNLO results are sensitive to the variation of the initial scale
$Q_0^2$. The values of $\Lambda_{\overline{MS}}^{(4)}$ become stable for
$Q_0^2\geq 10~GeV^2$ only. The same effect  manifests itself
for the results of the expanded [0/2] Pad\'e approximant fits, which
incorporate the explicit information about the N$^3$LO expression for the
coupling constant $A_s$. We think that this effect might be related
to a rather peculiar behaviour of the NNLO perturbative QCD expression
of $n=2$ moment. Indeed, taking into account the exact numerical
values of the coefficients $p(2)$, $q(2)$, $C^{(1)}(2)$ and $C^{(2)}(2)$
from Table 2, we find that the perturbative behaviour of this moment
is determined by the following perturbative series
\begin{equation}
AD(2,A_s)C_{NS}^{(2)}(A_s)=1-0.132A_s(Q^2)-46.155A_s^2(Q^2)+...
\end{equation}
where the relatively large $A_s^2$ coefficient is dominated by the
NNLO term of the coefficient function of $n=2$ moment.
Thus we think that it is more appropriate  to start the NNLO
QCD evolution from the
initial scale $Q_0^2=20~GeV^2$, where the numerical value of the
$A_s^2$ contribution in Eq.(29) is smaller. Note that this choice
of the initial scale  is also empirically
supported by the fact that it coincides with the middle  of $Q^2$ kinematic
region
of the CCFR data.

In Table 4 we  present the results of
our new fits
for $\Lambda^{(4)}_{\overline{MS}}$ and IRR-model parameter $A_2^{'}$
obtained in  LO,
NLO, NNLO and N$^3$LO ( modelled by the expanded and non-expanded
Pad\'e approximants) in
the cases of both
$\gamma\neq$0 and $\gamma=0$.

\begin{table}
%
%
\begin{tabular}{||c||c|c|c|c|c|c||} \hline
\hline
 &\multicolumn{3}{c|}{ $\gamma$ - free  } &\multicolumn{3}{c||}{ $\gamma = 0
$ - fixed  }  \\
\hline\hline
                $ Q^2 > $           &
                $\Lambda_{\overline{MS}}^{(4)}$ (MeV)     &
                $ A_2^\prime$(HT)   &
                $ \chi^2$/points    &
                $\Lambda_{\overline{MS}}^{(4)}$ (MeV)     &
                $ A_2^\prime$(HT) ($GeV^2)$)   &
                $ \chi^2$/points    \\
\hline\hline
 5 $GeV^2$ & &                 & & & &  \\
LO & 264$\pm$35      & -- & 113.1/86            &  241$\pm$35      & -- &
121.7/86              \\
   & 433$\pm$52      & -0.33$\pm$0.06 & 82.8/86 &  398$\pm$71      &
-0.31$\pm$0.08 & 121.7/86  \\
NLO & 339$\pm$36     & -- & 87.6/86             &  313$\pm$36      & -- &
95.2/86              \\
   & 369$\pm$45      & -0.12$\pm$0.06 & 82.3/86 &  341$\pm$36      &
-0.11$\pm$0.05 & 92.3/86  \\
NNLO & 326$\pm$35    & -- & 77.0/86             &  314$\pm$35      & -- &
86.1/86              \\
   & 326$\pm$35      & -0.01$\pm$0.05 & 76.9/86 &  315$\pm$34      &
-0.02$\pm$0.05 & 86.3/86  \\
N$^3$LO & 332$\pm$28 & -- & 76.9/86      &  314$\pm$28  & -- &
86.3/86     \\
(n.e.)      & 333$\pm$27 &  -0.04$\pm$0.05  & 76.3/86 & 315$\pm$27 &
 -0.04$\pm$0.05 & 85.7/86  \\
N$^3$LO & 335$\pm$37 & -- & 77.9/86             &  328$\pm$37      & -- &
85.1/86               \\
   & 340$\pm$37      & -0.04$\pm$0.05 & 77.2/86 &  335$\pm$37      &
 -0.05$\pm$0.05 & 84.2/86    \\
\hline
 10 $GeV^2$ & &                 & & & &  \\
LO & 287$\pm$42      & -- & 77.3/63             &  283$\pm$39      & -- &
78.1/63               \\
   & 529$\pm$77      & -0.52$\pm$0.12 & 57.8/63 &  515$\pm$75      &
-0.50$\pm$0.12 & 59.5/63   \\
NLO & 349$\pm$40     & -- & 63.9/63             &  344$\pm$44      & -- &
64.8/63               \\
   & 436$\pm$55      & -0.24$\pm$0.10 & 58.3/63 &  427$\pm$55      &
-0.23$\pm$0.10 & 59.5/63   \\
NNLO & 338$\pm$30    & -- & 57.4/63             &  337$\pm$40      & -- &
58.7/63               \\
  & 354$\pm$45       & -0.06$\pm$0.09 & 56.9/63 &  352$\pm$42      &
-0.03$\pm$0.09 & 58.2/63   \\
N$^3$LO & 348$\pm$41  & -- & 57.3/63             &  347$\pm$41   & -- &
58.0/63     \\
(n.e.)        & 373$\pm$48  & -0.09$\pm$0.09 & 56.2/63 & 373$\pm$48    &
-0.09$\pm$0.09 & 56.9/63  \\
N$^3$LO & 344$\pm$40 & -- & 56.8/63             &  345$\pm$30      & -- &
57.3/63               \\
     & 362$\pm$46    & -0.07$\pm$0.09 & 56.2/63 &  363$\pm$46      &
-0.07$\pm$0.09 & 56.6/63   \\
\hline
 15 $GeV^2$ & &                 & & & &  \\
LO & 319$\pm$48      & -- & 58.5/50             &  320$\pm$47      & -- &
58.5/50               \\
   & 530$\pm$89      & -0.56$\pm$0.18 & 49.9/50 &  525$\pm$45     &
-0.56$\pm$0.20 & 50.0/50   \\
NLO & 365$\pm$46     & -- & 52.3/50             &  366$\pm$46      & -- &
52.3/50               \\
   & 440$\pm$71      & -0.25$\pm$0.17 & 50.3/50 &  438$\pm$69      &
-0.24$\pm$0.17 & 50.3/50   \\
NNLO & 343$\pm$44    & -- & 50.4/50             &  341$\pm$44      & -- &
50.8/50               \\
   & 350$\pm$56      &  -0.03$\pm$0.15 & 50.3/50 &  347$\pm$55      &
-0.03$\pm$0.15 & 50.8/50   \\
N$^3$LO & 348$\pm$45  & -- & 50.3/50  & 348$\pm$45 & -- & 50.4/50 \\
(n.e.)        & 358$\pm$57 & -0.04$\pm$0.14 & 50.2/50 & 357$\pm$56 &
-0.04$\pm$0.15
    & 50.3/50 \\
N$^3$LO & 342$\pm$43 &    --            & 50.2/50 &  343$\pm$41      &
--
& 50.3/50   \\
   & 347$\pm$58      &  -0.02$\pm$0.15 & 50.2/50 &  347$\pm$54      &
-0.02$\pm$0.15 & 50.2/50   \\
\hline \hline
\end{tabular}
{{\bf Table 4}. The results of  extractions
of the parameter $\Lambda_{\overline{MS}}^{(4)}$ and
the IRR coefficient $A_2^{'}$, (in $GeV^2$) defined in Eq.(23), from LO,
NLO, NNLO
and N$^3$LO
non-expanded (n.e.) and expanded Pad\'e
fits of CCFR'97 data. In the fits we take $Q_0^2$=$20~GeV^2$}
\end{table}
\vspace{0.3cm}

Looking carefully at Table 4 we arrive at the following
conclusions:

\begin{itemize}

\item The results of  LO and NLO fits are  identical
to the ones obtained in Ref.\cite{KKPS2}.

\item Our fits demonstrate that
the NNLO values of the parameter $\Lambda_{\overline{MS}}^{(4)}$
depend on the choice of the initial scale $Q_0^2$.
In the case of  $Q_0^2=20~GeV^2$ the NNLO perturbative
QCD contributions
are less  important, than in the case $Q_0^2=5~GeV^2$,
earlier considered in Ref.\cite{KKPS2}.
Indeed, for different $Q^2$-cuts they  change slightly
the NLO values of $\Lambda_{\overline{MS}}^{(4)}$, provided
twist-4 corrections are switched off.
\item As was mentioned previously,  this effect might be related  to
the peculiar behaviour of the NNLO perturbative expression
for  $n=2$ moment of  $xF_3$  (see Eq.(29)) and, therefore,
to the theoretical uncertainty of the NNLO behaviour of
$xF_3$  at small $x$. We  checked this conclusion by comparing
the results of the NLO and NNLO $Q_0^2=5~GeV^2$-fits of the CCFR'97 data cut at
$x>0.04$.
The result of these test-fits demonstrate  the tendency, identical
to the one revealed after moving the initial scale to
$Q_0^2=20~GeV^2$, namely, the minimization of the difference
between the values of $\Lambda_{\overline{MS}}^{(4)}$, extracted
in NLO and NNLO.
\item
However, in the case when
the IRR-model for the twist-4 corrections are included
into the analysis, the effects of the NNLO corrections
are still important and  decrease both the value of
$\Lambda_{\overline{MS}}^{(4)}$ and the IRR-model parameter $A_2^{'}$,
making the first one almost identical to the NNLO value
of $\Lambda_{\overline{MS}}^{(4)}$, obtained without twist-4 terms.  
The latter one is compatible with zero within statistical
error bars. A similar feature was also observed
in the case of the fits, made in Ref.\cite{KKPS2} for
the initial scale $Q_0^2=5~GeV^2$.
This property  confirms the conclusion of Ref.\cite{KKPS2}
that the results
of the NNLO fits are less sensitive to the parameter
of the IRR-model of the twist-4 term. A similar conclusion was
also recently made in Ref.\cite{YB} while comparing
the experimental data for the DIS R-ratio with the available
NNLO perturbative QCD results of Ref.\cite{Neerven}, although
the earlier analysis of the experimental data for this
quantity with a different kind of the HT model  leaves still
room for the power suppressed behaviour \cite{santtw,KPSG}.

\item The values of $\Lambda_{\overline{MS}}^{(4)}$, obtained
from the fits  with the [0/2] Pad\'e
estimates (both in the expanded and non-expanded  variants)
turn out to be almost insensitive to the choice
of  the $Q^2$-cut
of the data,  the value of $\gamma$ and thus incorporation
of the $(1+\gamma x)$-factor in the parton distribution model.
The latter fact,
in  turn, can indicate that the change of
the  model
$xF_3(x,Q_0^2)=A(Q_0^2)x^{b(Q_0^2)}(1-x)^{c(Q_0^2)}
(1+\gamma(Q_0^2))x)$ to $xF_3(x,Q_0^2)=A(Q_0^2)x^{b(Q_0^2)}(1-x)^{c(Q_0^2)}
(1+\gamma(Q_0^2)x+\epsilon(Q_0^2)\sqrt{x})$, used in the
MRST and GRV fits, might  affect  the obtained results  only slightly;

\item Large errors in  definite results for
$\Lambda_{\overline{MS}}^{(4)}$, presented in Table 4,
reflect the correlations of these uncertainties with the errors
of the IRR-model parameter $A_2^{'}$;

\item The property $\chi^2_{LO}>\chi^2_{NLO}>\chi^2_{NNLO}\sim \chi^2_{N^3LO}$
reflects the importance of  the effects of
higher order perturbative QCD corrections in the process of  fits of
the concrete experimental data;

\item
For all $Q^2$-cuts the expanded N$^3$LO results for
$\Lambda_{\overline{MS}}^{(4)}$ are almost identical to the
ones obtained with
the expansion of the Pad\'e approximants in Taylor series.
Moreover, the $\chi^2$-criterion does not discriminate
between these two variants of the Pad\'e motivated fits
(see especially the results, obtained for the cuts
$Q^2>10~ GeV^2$ and $Q^2>15~GeV^2$).
In our future studies we will consider the results of
applications of the expanded Pad\'e approximants.
\end{itemize}

The results of the the NNLO fits, made with
the help of the Jacobi polynomial expansions method,  are
compared to the available CCFR'97 data in Fig.1. Drawing the
theoretical curves we used Eq.(24) with zero twist-4 contributions.
The value of the QCD scale parameter, which governs the theoretical
behaviour of the moments of $xF_3$ SF, turned out to be
$\Lambda_{\overline{MS}}^{(4)}=326\pm 35~MeV$ (the
$\chi^2$ of the corresponding fits is 77.0/86).
The values of the corresponding parameters of the $xF_3$  model
at $Q_0^2=20~GeV^2$  are : $A=4.70\pm 0.34$,
$b=0.65\pm 0.03$, $c=3.88\pm 0.08$, $\gamma=0.80\pm 0.28$.
One can see that the NNLO results of the fits without twist-4
corrections are in good agreement
with the CCFR'97 experimental data for  $xF_3$ .

To determine now the values of $\alpha_s(M_Z)$,
we transformed $\Lambda_{\overline{MS}}^{(4)}$
through the threshold of the production of the fifth
flavour, $M_5$. This is done
using the LO, NLO, NNLO and N$^3$LO variants of the
$\overline{MS}$-scheme matching conditions, derived in
Ref.\cite{CKS} following the lines of Ref.\cite{BW}.
The related values of $\Lambda^{(5)}_{\overline{MS}}$
can be obtained
with the help of  the following equation:


\newpage

\input epsf
\begin{figure}[t]
\centering
\hspace*{-5.5mm}
\leavevmode\epsfysize=6.5cm\epsfbox{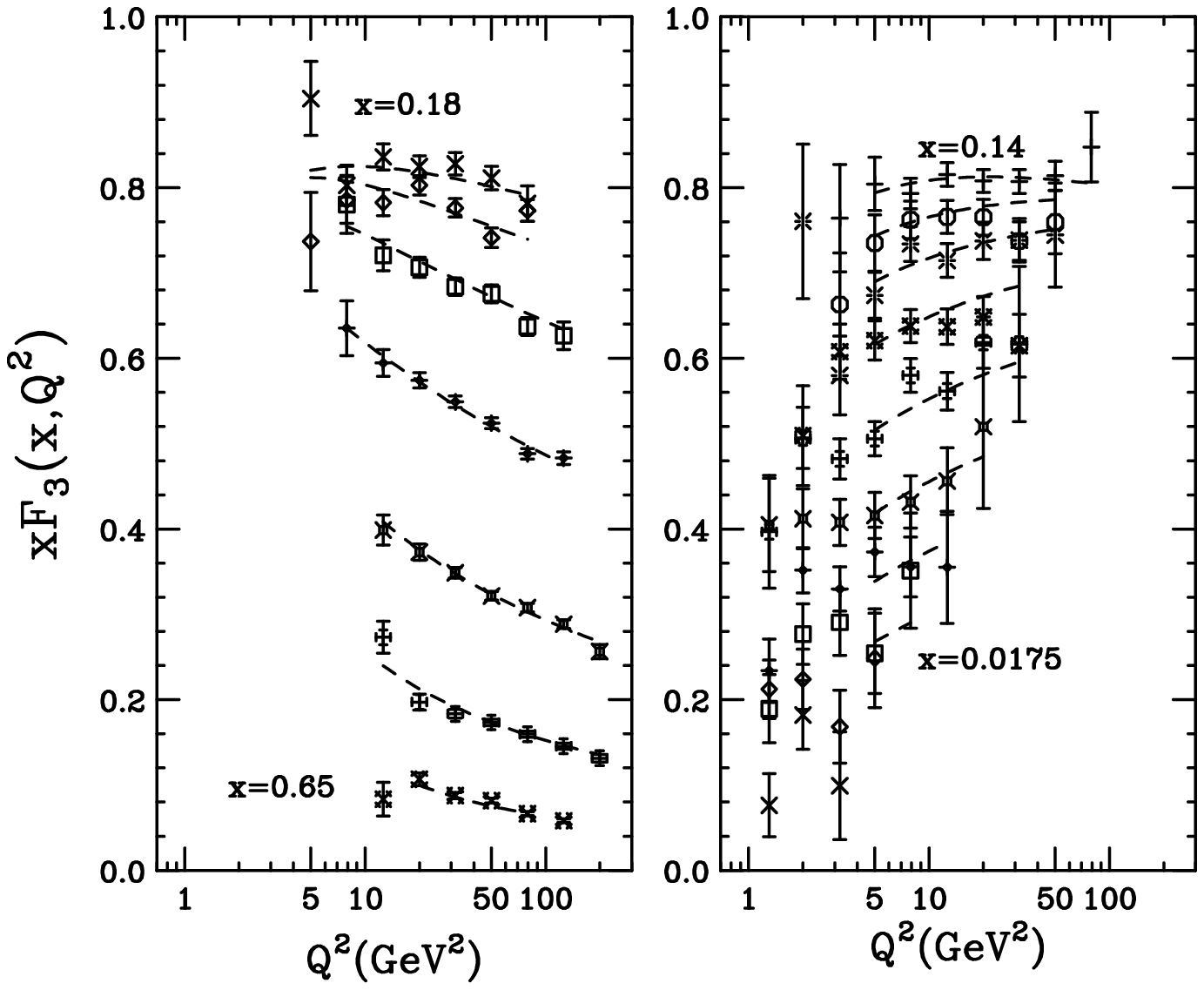}\\
\end{figure}
%
\begin{center}
Fig.1~The comparison of the CCFR'97 data with the results
of our NNLO Jacobi polynomial fits.
\end{center}

\begin{eqnarray}
\beta_0^{f+1}ln\frac{\Lambda
_{\overline{MS}}^{(f+1)~2}}{\Lambda_{\overline{MS}}^{(f)~2}}=
(\beta_0^{f+1}-\beta_0^f)L_h \\ \nonumber
+\delta_{NLO}+\delta_{NNLO}+
\delta_{N^3LO}
\end{eqnarray}
\begin{equation}
\delta_{NLO}=\bigg(\frac{\beta_1^{f+1}}{\beta_0^{f+1}}-\frac{\beta_1^f}
{\beta_0^f}\bigg)ln L_h-\frac{\beta_1^{f+1}}{\beta_0^{f+1}}ln
\frac{\beta_0^{f+1}}{\beta_0^f}
\end{equation}
\begin{eqnarray}
\delta_{NNLO}=\frac{1}{\beta_0^f L_h}\bigg[\frac{\beta_1^f}{\beta_0^f}
\bigg(\frac{\beta_1^{f+1}}{\beta_0^{f+1}}
-\frac{\beta_1^f}{\beta_0^f}\bigg)ln L_h
\\ \nonumber
+\bigg(\frac{\beta_1^{f+1}}{\beta_0^{f+1}}\bigg)^2
-\bigg(\frac{\beta_1^{f}}{\beta_0^{f}}\bigg)^2
-\frac{\beta_2^{f+1}}{\beta_0^{f+1}}
+\frac{\beta_2^{f}}{\beta_0^{f}}-C_2\bigg]
\end{eqnarray}
\begin{eqnarray}
\delta_{N^3LO}=\frac{1}{(\beta_0^f L_h)^2}\Bigg[
-\frac{1}{2}\bigg(\frac{\beta_1^f}{\beta_0^f}\bigg)^2
\bigg(\frac{\beta_1^{f+1}}{\beta_0^{f+1}}-\frac{\beta_1^f}{\beta_0^f}\bigg)
ln^2 L_h
\\ \nonumber
+\frac{\beta_1^f}{\beta_0^f}\bigg[-\frac{\beta_1^{f+1}}{\beta_0^{f+1}}
\bigg(\frac{\beta_1^{f+1}}{\beta_0^{f+1}}-\frac{\beta_1^f}{\beta_0^f}\bigg)
+\frac{\beta_2^{f+1}}{\beta_0^{f+1}}
-\frac{\beta_2^f}{\beta_0^f}+C_2\bigg]ln L_h
\\ \nonumber
+\frac{1}{2}\bigg(-\bigg(\frac{\beta_1^{f+1}}{\beta_0^{f+1}}\bigg)^3
-\bigg(\frac{\beta_1^f}{\beta_0^f}\bigg)^3-\frac{\beta_3^{f+1}}{\beta_0^{f+1}}
+\frac{\beta_3^f}{\beta_0^f}\bigg)
\\ \nonumber
+\frac{\beta_1^{f+1}}{\beta_0^{f+1}}\bigg(\bigg(
\frac{\beta_1^f}{\beta_0^f}\bigg)^2+
\frac{\beta_2^{f+1}}{\beta_0^{f+1}}-\frac{\beta_2^f}{\beta_0^f}+C_2\bigg)
-C_3\Bigg]
\end{eqnarray}
where $C_2=-7/24$ was calculated in Ref.\cite{LRVS}
(see also Erratum to Ref.\cite{BW}) and the analytic
expression for $C_3$, namely  $C_3=-(80507/27648)\zeta(3)-(2/3)\zeta(2)
((1/3) ln2+1)-58933/124416 +(f/9)[\zeta(2)+2479/3456]$ was recently found
in Ref.\cite{CKS}. Here
$\beta_i^f$ ($\beta_i^{f+1}$) are  coefficients of the
$\beta$-function with $f$ ($f+1$)
numbers of active flavours,
$L_h=ln(M_{f+1}^2/\Lambda_{\overline{MS}}^{(f)~2})$ and $M_{f+1}$ is the
threshold of the production of a quark of the $(f+1)$ flavour.
In our analysis we will take $f=4$ and $m_b\approx 4.8~GeV$
and  vary the threshold of the production of the fifth
flavour from $M_5^2=m_b^2$ to $M_5^2=(6m_b)^2$ in accordance
with the proposal of Ref.\cite{BN}.
The difference between different prescriptions of fixing
the matching point is included into the estimate of the
theoretical uncertainties of the final results for $\alpha_s(M_Z)$.

In the case of the non-zero values of the twist-4 function
$h(x)\neq 0$, the results of the fits are presented
in Table 5
in the next Section.

It should be stressed that we  consider the outcomes of our N$^3$LO
approximate fits as  theoretical uncertainties  of the
NNLO results in the same manner as the results of the NNLO analysis
are considered as the measure of theoretical uncertainties
of the NLO results.
In particular, we  introduce the characteristic deviations
$\Delta^{NNLO}$=
$|(\Lambda_{\overline{MS}}^{(4)})^{N^3LO}-
(\Lambda_{\overline{MS}}^{(4)})^{NNLO}|$,
$\Delta^{NLO}$=
$|(\Lambda_{\overline{MS}}^{(4)})^{NNLO}-
(\Lambda_{\overline{MS}}^{(4)})^{NLO}|$.

When the twist-4 terms are included into the fits,
the difference
$\Delta^{NNLO}$=
$|(\Lambda_{\overline{MS}}^{(4)})^{N^3LO}-
(\Lambda_{\overline{MS}}^{(4)})^{NNLO}|$
is  smaller than the NLO correction term
$\Delta^{NLO}$=
$|(\Lambda_{\overline{MS}}^{(4)})^{NNLO}-
(\Lambda_{\overline{MS}}^{(4)})^{NLO}|$. A similar tendency
$\Delta^{NNLO}<\Delta^{NLO}$  takes place in the case
of the fits without twist-4 corrections.
These observed properties demonstrate
the reduction of  theoretical errors due to cutting
the analyzed perturbation series in different orders.

It is known that the inclusion of  higher-order perturbative
QCD corrections in
the  comparison with
experimental data  decreases the scale-scheme theoretical
errors of the results for $\Lambda_{\overline{MS}}^{(4)}$ and thus
$\alpha_s(M_Z)$ (see e.g. Refs.\cite{ChK,PKK,EGKS}). Among the ways of
probing the scale-scheme uncertainties are the scheme-invariant
methods, namely the principle of minimal sensitivity,
the effective charge approach
( which is known to be
identical to the  scheme-invariant perturbation theory) and the
BLM approach (for a review of these methods see e.g. Ref.\cite{ChK2}).
The scheme-invariant methods
were already used to estimate  the
unknown higher order corrections in SFs (see Ref.\cite{KPSG}, where
a strong decrease of the value of the QCD scale
parameter was found in  the NLO scheme-invariant
fit of the experimental data for the NS part of the $F_2$  ). It was also
used  to
try to
estimate the unknown
at present N$^3$LO corrections to  definite physical
quantities \cite{KatSt}, and DIS sum rules among others.
Note that the predictions of Ref.\cite{KatSt} turned out
to be  in agreement
with the results of applications of the Pad\'e resummation technique
(see Ref.\cite{SEK}). Therefore, we can conclude that the application
of the methods of the Pad\'e approximants should  lead to the reduction
of the scale-scheme dependence uncertainties of the values of
$\alpha_s(M_Z)$ in the analysis of the CCFR data.

To consider the applicability of the
Pad\'e resummation technique for fixing scale-scheme dependence
ambiguities, we  performed the scheme-invariant fits following
the ideas of Ref.\cite{Grun1}. We  found that in NNLO
the application
of the effective charge approach gave a rather high value of
$\chi^2$ ($\chi^2 \sim$ 111/86). This, in a turn, can be related to the
appearance of   $\underline{large}$ and $\underline{positive}$
values of the NNLO terms $\beta_2(n)_{eff}$ of the effective-charges
$\beta$-functions,  which are
the important ingredients of the scheme-invariant approach of Ref.\cite{Grun1}.
Similar problems have also been  observed
in the case of the scheme-invariant
applications to the study of the NNLO perturbative QCD predictions
to other renormalization-group invariant quantities
(see Refs.\cite{GKLS,gam2,ChKL} for  discussions). In this work
we  avoid the detailed investigation of this problem.
However, in Sec.4 we will discuss the results, which were obtained
after variation of the factorization and renormalization scales without
using the freedom in the choice of the scheme-dependent
NLO and NNLO corrections to
the anomalous dimension functions and  NNLO corrections
to the QCD $\beta$-function.


We now present  the values of $\alpha_s(M_Z)$, extracted
from the fits of the CCFR'97 experimental data
for  $xF_3$, obtained with the  twist-4
contribution, modelled  through the IRR model of Ref.\cite{DW}:

\begin{eqnarray}
NLO~HT~of~ Ref.\cite{DW}~\alpha_s(M_Z) =0.120 \pm 0.003(stat)
\\ \nonumber
\pm 0.005(syst)
\pm 0.004(theory)
\end{eqnarray}
\begin{eqnarray}
NNLO~HT~of~ Ref.\cite{DW}~\alpha_s(M_Z)=0.118\pm 0.002(stat)
\\ \nonumber
\pm 0.005(syst)
\pm 0.003(theory)~.
\end{eqnarray}

Anticipating  the considerations of Sec.3(b) we also present
the results of  NLO and NNLO extractions of
$\alpha_s(M_Z)$ with
the twist-4 contribution, modelled by
additional free parameters of the fit:

\begin{eqnarray}
NLO~HT~free~\alpha_s(M_Z) =0.123^{+0.008}_{-0.010}(stat)
\\ \nonumber
\pm 0.005(syst)
\pm 0.004(theory)
\end{eqnarray}
\begin{eqnarray}
NNLO~HT~free~\alpha_s(M_Z)=0.121^{+0.007}_{-0.009}(stat)
\\ \nonumber
\pm 0.005(syst)
\pm 0.003(theory)~~~.
\end{eqnarray}
Systematic uncertainties are taken from the
CCFR experimental
analysis, presented in the first work of Ref.\cite{Seligman},
and the theoretical uncertainties in the results
of Eqs.(34),(36) [Eqs.(35),(37)] are estimated by  differences
between
central values of the outcomes of the NNLO and NLO [N$^3$LO and NNLO] fits,
presented in Tables 4,5, plus  in the application
of
the $\overline{MS}$-scheme
matching condition (which following the
considerations of Ref.\cite{BN} was estimated as $\Delta\alpha_s(M_Z)=
\pm 0.002$).
In the process of fixing   the theoretical
errors with the consideration of the N$^3$LO corrections we take into account
the differences between  applications of the expanded and non-expanded
Pad\'e approximants.

It can be seen that due to a large overall number of the fitted
parameters the results of Eqs.(36),(37) for
$\alpha_s(M_Z)$ obtain  large statistical
uncertainties. As can be seen from the results of Eq.(34),(35)
for the QCD coupling constant,
it is possible to decrease their values by fixing the
concrete form of the twist-4 parameter $h(x)$.
However, if one is interested in the extraction of the form of the twist-4
parameter $h(x)$, one should take for granted these intrinsic
theoretical uncertainties of the value of $\alpha_s(M_Z)$.

{\bf 3 (b). The  analysis of the experimental data:
the extraction of the shape  of the twist-4 terms.}

Apart from the perturbative QCD contributions, the  expressions
for DIS structure functions should contain  power-suppressed
high-twist terms, which reflect  possible non-perturbative
QCD effects. The  studies of these terms have  a rather
long history. At the beginning of these studies it was
realized that the twist-4 contributions to structure functions
should have the pole-like behavior $\sim 1/((1-x)Q^2)$
\cite{Stw,Nason}. This behavior was used in the
phenomenological
investigations of the earlier less precise  DIS $\nu N$ data
\cite{Barnett,Isaev,f3jac}, which
together with  other different  procedures
of analyzing neutrino DIS data \cite{DeRujula,Yndurain}
was considered as the source of the information about
scaling violation parameters. The development of the renormalon
technique (See Refs.\cite{IRR,DW,MSSM} and Ref.\cite{Beneke} for a
detailed review) pushed ahead a more detailed phenomenological
analysis of the possibility of detecting higher-twist components
in the  most precise DIS data available at present, obtained by
BCDMS, SLAC, CCFR and other collaborations.
It turned out that despite the qualitative status of the renormalon
approach, a satisfactory description of the results of the QCD
NLO $F_2$  analysis \cite{VM}
in terms of the IRR technique was achieved
\cite{DW,IRR}.
The next step was to clarify the status of the predictions
of Ref.\cite{DW} for the form and sign of the twist-4 contributions
to $xF_3$. The study of this problem was done in
Ref.\cite{KKPS2} (see also Ref.\cite{S}).
In this section, we  discuss the results of a more refined
analysis of the behavior of the twist-4 contributions  to
$xF_3$  in LO, NLO, NNLO and beyond.

In Table 4 we present the dependence of the extracted value
of the  parameter $A_2^{'}$ on  different orders of
perturbative QCD predictions,
$Q^2$-cuts of the CCFR experimental data and the coefficient $\gamma$
of the parton distribution model for $xF_3$, fixing the factorization scale
$Q_0^2=20~GeV^2$. Note
that
the parameter $A_2^{'}$  was introduced in the IRR model
of Eq.(28), taken from Ref.\cite{DW}, and fixed there as $A_2^{'}\approx
-0.2~GeV^2$, which is necessary for the description of the
fitted twist-4 results of Ref.\cite{VM} for  $F_2$  in the
IRR language. We found that the value of this parameter,
extracted in  LO and NLO, is negative, differs from zero for
about one standard deviation and qualitatively agrees with the
IRR-motivated guess of Ref.\cite{DW}. Moreover, the results of our
LO and NLO fits
are also in agreement with
the value of the parameter   $h=-0.38\pm 0.06~GeV^2$ of a different
model of the
twist-4 contribution to $xF_3$, namely $xF_3(x,Q^2)h/((1-x)Q^2)$,
extracted previously in Ref.\cite{f3jac} from the old $\nu N$ DIS data.

It is interesting to notice  that the results of Table 4
reveal that for larger $Q^2$-cuts $10-15~GeV^2$ the values of
$A_2^{'}$ in the LO and NLO fits are less sensitive to the included
number of  experimental points
than in the case of the  low
$Q^2$ cut ($5~GeV^2$). This feature can be related to the logarithmic
increase of the QCD coupling constant $A_s$ at lower $Q^2$.
However, since we are interested in the extraction of the
power-suppressed twist-4 contribution, we  concentrate
on the discussion of  more informative, from our point of view,
fits with low $Q^2$-cut $5~GeV^2$, which contain more experimental points
and thus are  more statistically motivated.

We now turn to a pure phenomenological
extraction of the twist-4 contribution $h(x)$ to
$xF_3$ (see Eq.(24)), which is motivated by the
work of Ref.\cite{VM} for  $F_2$. In the framework
of this approach the $x$-shape of $h(x)$ is parametrized by
additional parameters $h_i=h(x_i)$, where $x_i$ are
points of the experimental
data bining. The results of the multiloop extractions of these
parameters are presented in Table 5 and are illustrated in Fig.2.

\begin{center}
\begin{tabular}{||c|c|c|c|c||} \hline \hline
                    &~~      LO            &       NLO       &   NNLO &
N$^3$LO
(Pade)  \\ \hline
 $\chi^2_{d.f.}$    &  66.3/86           &        65.7/86  &  65.0/86 &
64.8/86          \\
  A    &   5.33 $\pm$  1.33  &  4.71 $\pm$  1.14   &  4.79 $\pm$ 0.75 &
5.14 $\pm$ 0.73  \\
  b    &   0.69 $\pm$  0.08  &  0.66 $\pm$  0.08   &  0.66 $\pm$ 0.05 &
0.68 $\pm$ 0.05  \\
  c    &   4.21 $\pm$  0.17  &  4.09 $\pm$  0.14   &  3.95 $\pm$ 0.19 &
3.84 $\pm$ 0.23  \\
$\gamma$  &1.15 $\pm$  0.94  &  1.34 $\pm$  0.86   &  0.96 $\pm$ 0.57 &
0.57 $\pm$ 0.52  \\
$\Lambda_{\overline{MS}}^{(4)}$ $[MeV]$ & 331 $\pm$ 162 & 440 $\pm$ 183  &
372 $\pm$ 133  & 371 $\pm$ 127\\ \hline \hline
   $x_i$                     &\multicolumn{4}{c||}{  $h(x_i)~[GeV^2]$ }
\\  \hline
0.0125 &   0.209 $\pm$ 0.346 &    0.235  $\pm$ 0.325 &     0.263 $\pm$
0.315 & 0.304  $\pm$ 0.313  \\
0.0175 &   0.067 $\pm$ 0.281 &    0.114  $\pm$ 0.283 &     0.133 $\pm$
0.263 & 0.170  $\pm$ 0.257 \\
0.025  &   0.153 $\pm$ 0.215 &    0.242  $\pm$ 0.226 &     0.244 $\pm$
0.206 & 0.268  $\pm$ 0.199 \\
0.035  &  -0.013 $\pm$ 0.205 &    0.132  $\pm$ 0.236 &     0.112 $\pm$
0.211 & 0.110  $\pm$ 0.195 \\
0.050  &   0.038 $\pm$ 0.159 &    0.256  $\pm$ 0.240 &     0.214 $\pm$
0.200 & 0.171  $\pm$ 0.168 \\
0.070  &  -0.141 $\pm$ 0.139 &    0.144  $\pm$ 0.258 &     0.106 $\pm$
0.202 & 0.017  $\pm$ 0.151 \\
0.090  &  -0.177 $\pm$ 0.127 &    0.144  $\pm$ 0.254 &     0.142 $\pm$
0.202 & 0.026  $\pm$ 0.142 \\
0.110  &  -0.343 $\pm$ 0.127 &   -0.009  $\pm$ 0.237 &     0.045 $\pm$
0.205 & -0.080 $\pm$ 0.146 \\
0.140  &  -0.408 $\pm$ 0.116 &   -0.085  $\pm$ 0.174 &     0.060 $\pm$
0.187 & -0.049 $\pm$ 0.140 \\
0.180  &  -0.351 $\pm$ 0.173 &   -0.072  $\pm$ 0.128 &     0.154 $\pm$
0.161 &  0.093 $\pm$ 0.145 \\
0.225  &  -0.547 $\pm$ 0.244 &   -0.351  $\pm$ 0.180 &    -0.098 $\pm$
0.128 & -0.103 $\pm$ 0.127 \\
0.275  &  -0.548 $\pm$ 0.334 &   -0.472  $\pm$ 0.321 &    -0.228 $\pm$
0.185 & -0.193 $\pm$ 0.169 \\
0.350  &  -0.295 $\pm$ 0.414 &   -0.390  $\pm$ 0.470 &    -0.154 $\pm$
0.264 & -0.104 $\pm$ 0.234 \\
0.450  &  -0.098 $\pm$ 0.410 &   -0.307  $\pm$ 0.504 &    -0.146 $\pm$
0.330 & -0.125 $\pm$ 0.303 \\
0.550  &   0.095 $\pm$ 0.324 &   -0.134  $\pm$ 0.415 &    -0.121 $\pm$
0.323 & -0.138 $\pm$ 0.317 \\
0.650  &   0.380 $\pm$ 0.211 &    0.215  $\pm$ 0.262 &     0.140 $\pm$
0.245 & 0.111  $\pm$ 0.251 \\
\hline  \hline
\multicolumn{5}{p{15cm}}{{\bf Table 5.}
The results of
extractions of the HT contribution $h(x)$ to $xF_3$ and
the parameters $A,b,c,\gamma$ with the corresponding
statistical errors. The
QCD fits of  CCFR'97 data were performed  taking
into account TMC in LO, NLO
($N_{max}=10$), NNLO and N$^3$LO ($N_{max}=6$).
In the latter case the
expanded [0/2]  Pad\'e approximants were used.
The fits are done for $Q_0^2=20$ $GeV^2$.}\\
\end{tabular}
\end{center}

\newpage
\input epsf
\begin{figure}[t]
\centering
\hspace*{-5.5mm}
\leavevmode\epsfysize=10cm\epsfbox{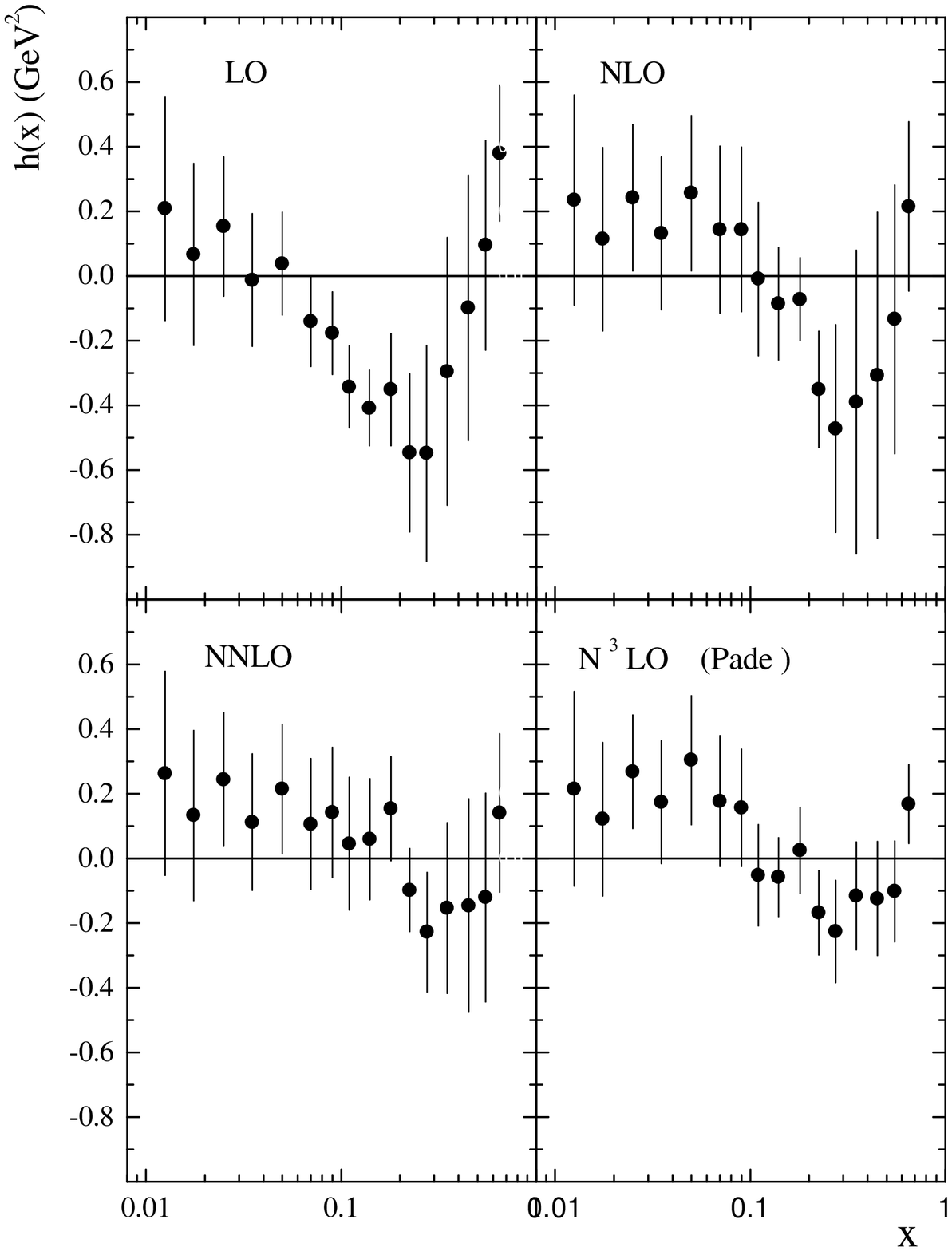}
\end{figure}
\begin{center}
Fig.2. The $x$-shape of $h(x)$ extracted from the fits of
CCFR'97 data in the case of fixing  the intial
scale at $Q_0^2=20~GeV^2$
\end{center}

Looking carefully at Table 5 and Fig.2 we observe the following features:
\begin{enumerate}
\item The $x$-shape of the twist-4 parameter
is not inconsistent with the
expected rise of
$h(x)$ for $x\rightarrow 1$ \cite{Stw,Nason} in all orders of perturbation
theory;
\item the values of  parameters $h(x_i)$ at the upper and lower
points of the kinematic region ($x_{16}$=0.650 and $x_{1}$=0.0125)
are stable to the inclusion of  higher order perturbative QCD corrections
and application of the Pad\'e resummation technique. At large values of
$x$ this feature is in agreement with the previous statement;
\item the function  $h(x)$ seems to cross zero twice:
at small $x$ of order 0.03
and larger $x$ about 0.4. It should be noted that the sign-alternating
behavior of the twist-4 contributions to DIS structure functions
was qualitatively predicted in Ref.\cite{Isaev};
\item in  LO and NLO our results are in qualitative agreement
with the IRR prediction of Ref.\cite{DW} (for discussions, see
Ref.\cite{Beneke});
\item in  NNLO this agreement is not so obvious, though a certain
indication to the manifestation of
the twist-4 term  survives even
in NNLO;



\item however, in NNLO we observe  the
minimization of the amplitude of the $h(x)$-variation.
Thus we conclude that the
inclusion of the NNLO corrections into the game might "shadow" the
effects of the power suppressed terms
in NNLO.
This property was previously
observed in LO as a result of the analysis of  less precise DIS neutrino
data in Ref.\cite{Barnett}. In the modern experimental situation, namely,
in  the analysis of  more precise DIS neutrino data
of the CCFR collaboration, we  observe this feature in NNLO;
\item we checked the reliability of this NNLO foundation
by going
beyond this
perturbative approximation using the method of Pad\'e approximants.
The result of this analysis reveals a relative stability of the NNLO results
for $h(x)$;
\item the property of  minimization of the $x$-shape  of $h(x)$
in NNLO and N$^3$LO is identical to the effect
of decreasing   the IRR model parameter $A_2^{'}$
in NNLO and N$^3$LO (see Table 4);
\item these observed properties clarify why the results of the NNLO
and N$^3$LO fits for $\Lambda_{\overline{MS}}^{(4)}$ , presented in Tables 4,5
practically
do not depend  on the inclusion of the twist-4 contribution
using the IRR model.
Indeed, at this level the twist-4 terms  manifest themselves
less obviously.
\end{enumerate}

From our point of view,    statements (8)-(9)
reflect  self-consistency of the results of our different
fits with twist-4 terms included in different ways.

{\bf 4. The  factorization and renormalization
scales uncertainties.}

As it is known from the work of Ref.\cite{VM} 
it is rather instructive to study the sensitivity of the
results to the
variation of renormalization and factorization scales.
We will  study the question of factorization-renormalization
scale dependence within the class of $\overline{MS}$-scheme only.
This means that we will change
 only the scales without varying the scheme-dependent
coefficients of anomalous dimensions
beta-function.

The  arbitrary factorization scale is entering in the following
equation:
\begin{equation}
A_s( Q^2/\mu^2_{\overline{MS}})= A_s(Q^2/\mu_F^2)\bigg[1+k_1 A_s(Q^2/\mu_F^2)+
k_2 A_s^2(Q^2/\mu_F^2)  \bigg]
\end{equation}
where $\mu_F^2$  is the factorization scale
and
\begin{eqnarray}
k_1&=&-\beta_0 ln(\frac{\mu^2_{\overline{MS}}}{\mu_F^2}) \\ \nonumber
k_2&=&k_1^2+\frac{\beta_1}{\beta_0}k_1
\end{eqnarray}
Let us choose the factorization scale as $\mu_F^2=\mu_{\overline{MS}}^2 k_F$.

Then we have:
\begin{equation}
k_1=\beta_0 ln(k_F)
\end{equation}

In this case after application of the renormalization group
equation and substitution of Eq.(38) into Eqs.(7,8) of Sec.2
we get
\begin{equation}
exp\bigg[-\int^{A_s(Q^2 k_F)}\frac{\gamma_{NS}^{(n)}(x)}{\beta(x)}dx\bigg]
=(A_s(Q^2k_F))^{a}\times \overline{AD}(n,A_s(Q^2 k_F))
\end{equation}
where $a=\gamma_{NS}^{(0)}/2\beta_0$ and
\begin{eqnarray}
\overline{AD}(n,A_s(Q^2 k_F))&=&1+\bigg[p(n)+a k_1\bigg]A_s(Q^2 k_F) \\ \nonumber
&+&\bigg[q(n)+p(n)k_1(a+1)+\frac{\beta_1}{\beta_0}k_1
a+\frac{a(a+1)}{2}k_1^2\bigg]A_s^2(Q^2 k_F)
\end{eqnarray}
where
\begin{equation}
k_1=  ln(k_F)\beta_0
\end{equation}

Now let us study the factorization and renormalization scale dependence 
in the  case when $k_R=k_F=k$ (see, e.g., Ref.\cite{VV}).
In this case we should modify the coefficient function in Eq.(6) only as
\begin{equation}
C_{NS}^{(n)}=1+C^{(1)}(n)A_s(Q^2 k)+\bigg[C^{(2)}(n)+
C^{(1)}(n)\beta_0 ln(k)
\bigg]A_s^2(Q^2 k)
\end{equation}

The commonly accepted practice is  vary $k$
in the interval $1/4\leq k\leq 4$ (see, e.g., Ref.\cite{VV})

We repeated our fits, described in Sec.3a, both without
and with IRR model of the twist-4 terms
in the cases of $k=1/4$ and $k=4$.
The  obtained results are presented in Table 6:

\begin{center}
\begin{tabular}{||r|r|r|c|c|}
\hline
   Order &  k & $\Delta_k$      & $A'_2~(HT)$  &  $\chi^2$/points       \\
\hline
   NLO   &   4    &   116       &       ---    &    99.1/86       \\
         &   4    &   213       &    - 0.22$\pm$0.06  &    84.2/86       \\
         &   1/4  &   -61       &       ---       &    80.4/86       \\
         &   1/4  &   -99       &    + 0.02$\pm$0.05  &    80.2/86       \\
\hline   &        &             &                  &                  \\
   NNLO  &   4    &   35        &       ---       &    83.5/86       \\
         &   4    &   66        &    - 0.11$\pm$0.06  &    83.5/86       \\
         &   1/4  &   -51       &       ---       &    87.3/86       \\
         &   1/4  &   -45       &    + 0.09$\pm$0.05  &    84.5/86       \\
\hline
\end{tabular}
\end{center}
{{\bf Table 6.}
The results of NLO and NNLO fits of CCFR'97 $xF_3$ data for $Q^2>5~GeV^2$ with different values
of factorization/renormalization scales.
$\Delta_k (MeV)=\Lambda^{(4)}_{\overline{MS}}(k)
-\Lambda^{(4)}_{\overline{MS}}(k=1)$. The value of the
IRR model coefficient is given in $GeV^2$. 
The initial scale is fixed as $Q_0^2=20~GeV^2$.}

\newpage

The factorization/renormalization scale ambiguities
of the NLO (NNLO) values of $\Lambda_{\overline{MS}}^{(4)}$
are giving the estimate
of theoretical errors of $\alpha_s(M_Z)$, which can be
compared with the ones, given in Sec.3 (see Eqs.(34),(35)).
Transforming the results of Table 6 into the related errors 
of $\alpha_s(M_Z)$  with the help of NLO 
and NNLO variants of the matching condition of Eq.(30) 
we get:
\begin{eqnarray}
\Delta \alpha_s(M_Z)_{NLO}  &=&~^{+0.009}_{-0.007}   \\ \nonumber
\Delta \alpha_s(M_Z)_{NNLO} &=&~^{+0.003}_{-0.002}
\end{eqnarray}
in the case when HT are included through the IRR model
and
\begin{eqnarray}
\Delta \alpha_s(M_Z)_{NLO}  &=&~^{+0.006}_{-0.004}   \\ \nonumber
\Delta \alpha_s(M_Z)_{NNLO} &=&~^{+0.003}_{-0.002}
\end{eqnarray}
in the case when HT-terms are neglected.

One can see,  that the inclusion of the HT terms
even through the definite model leads to the increase of
theoretical uncertainties of the NLO value of the $\alpha_s(M_Z)$.
However, for the value of $k=1/4$ the fitted value 
of $A_2'(HT)$ is lying closer to zero.
This feature demonstrates once more that the inclusion of 
the NNLO corrections, modeled using the renormalization/factorization 
scale dependence have the tendency to decrease the fitted value 
of IRR model parameter. Another interesting observation is that 
at the NNLO the scale-dependence is drastically smaller.

It should be stressed, that 
the results obtained without HT terms are in
agreement with the results of the NLO and NNLO NS analysis of 
theoretical predictions
for the NS part of $F_2$ SF, which were obtained  recently in Ref.\cite{VV}
with the help of DGLAP approach. Moreover, considering the case 
when $k_F\neq k_R$ and varying both parameters within 
the interval $1/4 \leq k_F\leq 4$ and $1/4\leq k_R\leq 4$, 
we found that in NNLO the results of Table 6 depend only slightly  
on the choice of the factorization scale and are mostly 
related to the different values of renormalization scale.
This feature confirms our findings that the  NNLO corrections 
to the coefficient functions are more important 
than the corrections of the same order to the anomalous dimensions 
functions.

The next wellcomed feature is that 
the renormalization/factorization- scale estimate of  
theoretical uncertainty of the final NNLO
results are in agreement with the theoretical uncertainty of the 
NNLO result of Eq.(35), estimated by means of application  
of the Pad\'e approximation technique at the N$^3$LO.
This fact gives additional theoretical support in favour 
of the reliability of the NNLO value of $\alpha_s(M_Z)$, given in Eq.(35), 
which is one of the main results of our considerations.  

Let us discuss the renormalization/factorization-scale 
dependence of the results of the fits,   
performed in the case when   HT terms were included as the free 
parameters and $\gamma=0$ (for simplicity).  In this case we  found the following scale-dependence 
of the central values of $\alpha_s(M_Z)$, given in Eqs.(36),(37):
\begin{eqnarray}
\Delta \alpha_s(M_Z)_{NLO}  &=&~^{+0.006}_{-0.005}   \\ \nonumber
\Delta\alpha_s(M_Z)_{NNLO} &=&~\pm {0.003}
\end{eqnarray}

\input epsf
\begin{figure}[t]
\centering
\hspace*{-5.5mm}
\leavevmode\epsfysize=10cm\epsfbox{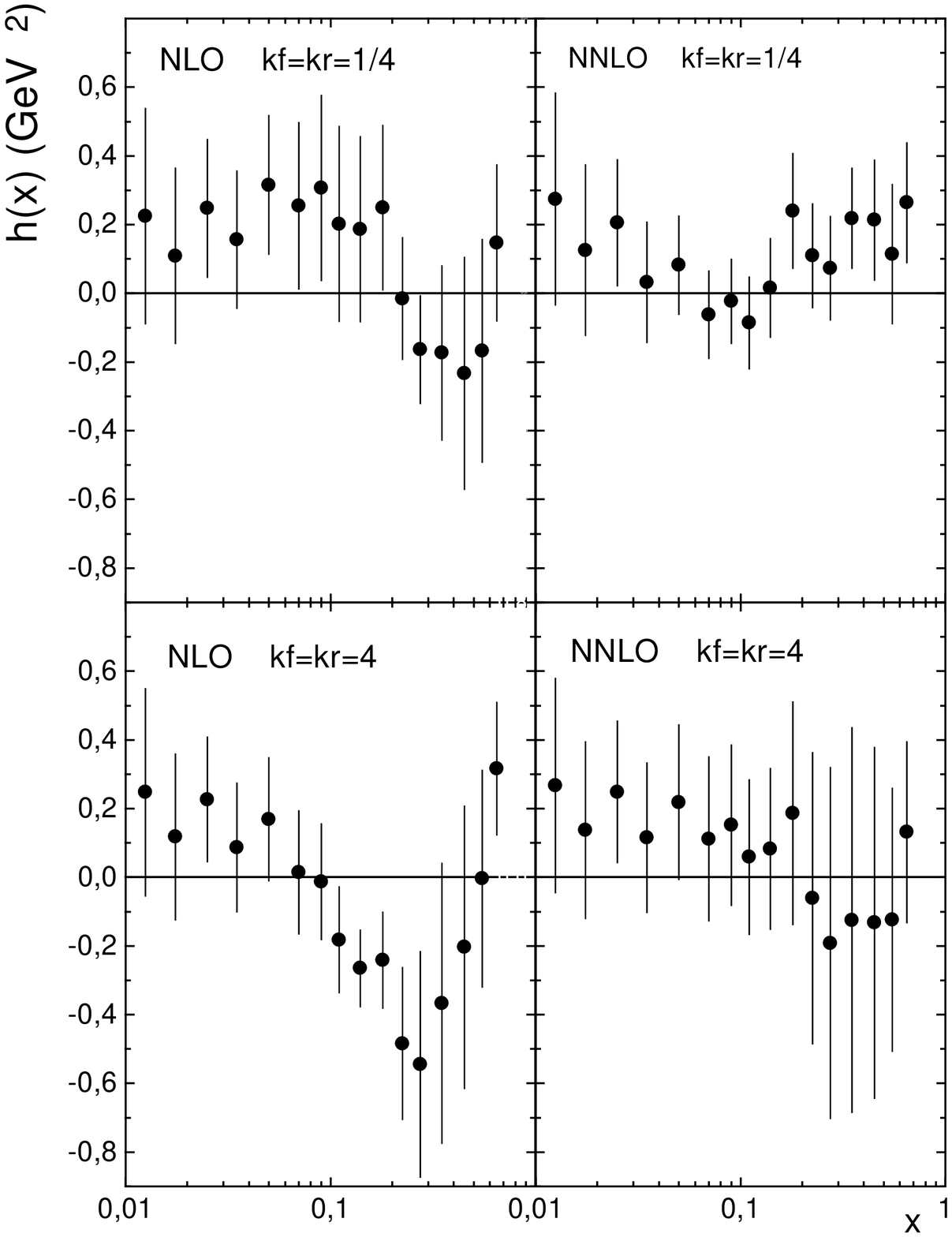}
\end{figure}
\begin{center}
Fig.3.~ 
The $x$-shape of $h(x)$ extracted from the fits of
CCFR'97 data in the case of fixing  the initial
scale at $Q_0^2=20~GeV^2$ and changing factorization/renormalization 
scales.
\end{center}

These errors  demostrate the same feautures, as 
in the case of the inclusion of the HT-terms through the 
IRR-model, namely 

1) the reduction of theoretical errors in NNLO;

2) the  correspondence of the theoretical error in  NNLO 
to the one, previously estimated with the 
help of simulation of the effects of the N$^3$LO corrections 
by means of the  
Pad\'e resummation technique;

3) the agreement of the results of Eq.(47) with the 
independent estimates of factorization/renormalization 
scale uncertainties, recently obtained  in the NS case 
in Ref.\cite{VV}.

In Fig.3 we present the investigation of the scale-dependence of the 
$x$-shape of the  model-independent twist-4 coefficient  $h(x)$,
which was obtained in Sec.3(b)   in LO,  NLO, NNLO and 
Pad\'e-motivated N$^3$LO in the cases of $k_F=k_R=1$ (see Fig.2).
More definitely, Fig.3 demonstrate, how  the NLO and NNLO plots of  Fig.2 
are changing in two cases, namely when  $k_F=k_R=1/4$ 
and $k_R=k_F=4$.

These plots demonstrate us the following typical feautures 
of scale-dependence of the $x$-shape of the twist-4 contributions:

\begin{itemize}

\item In the case of $k_R=k_F=4$ the result of the NLO analysis 
is closer to the LO behaviour of $h(x)$, obtained in Fig.2 
in the case of $k_R=k_F=1$, while in the case of 
$k_R=k_F=1/4$ the NLO $x$-shape of the twist-4 correction 
is almost identical to the one, obtained in Sec.3(b) in NNLO 
and indicate the partial reduction of the twist-4 model-independent 
contributions. This means, that  considering the 
renormalizatin/factorization scale ambiguities we can simulate in part 
the effects of the NNLO contributions. This feature confirmes 
the findings of Ref.\cite{AK2}, obtained at the NLO within 
DGLAP approach. 

\item In the case of $k_R=k_F=1/4$ both NLO and  NNLO $x$-shapes of $h(x)$ 
are in  agreement with the results of applications 
of the N$^3$LO Pad\'e approximation technique, shown 
in Fig.2. 

\item In the case of $k_R=k_F=4$ the NNLO $x$-shape of $h(x)$ 
is also comparable with minimization of the contribution 
of the twist-4 corrections. However, for these values 
of renormalziation/factorization scales  
the detailed structure 
of $h(x)$ is less vivid and has large errors.
This property is explained by the effective enhancement 
of the contributions of the NNLO corrections to the 
coefficient functions of the corresponding Mellin moments 
(see Table 2 and Eq.(44)) and large correlations 
of HT effets with the values of the parameter 
$\Lambda_{\overline{MS}}^{(4)}$.  

\end{itemize}

{\bf 5. The attempt of  the inclusion  of nuclear
corrections.}

The effects of nuclear corrections
are the remaining  important source of uncertainties of
the analysis of the DIS data.
This  is especially important for the experiments
on heavy targets and in the case of CCFR data--on iron  $^{56}Fe$.

The attempts to study these effects were done in Ref.\cite{SidTok}
in the framework of the Deuteron-motivated model. A
satisfactory QCD description of the CCFR data for $xF_3$ was achieved
due to the fact that in this case the nuclear effects
do not exceed 5 $\%$ effect. However, a more
realistic description of the nuclear effect for  $xF_3$
in the case of  $^{56}Fe$-target \cite{Kulagin}
revealed the appearance of new $1/Q^2$ and $1/M$ corrections
to   NS moments,
which have the following form
\begin{eqnarray}
M_{n}^{A}(Q^2)/A&=&\bigg(1+\frac{\epsilon}{M}(n-1)+\frac{<\bf{p}^2>}{6M^2}n(
n-1)
+ O(\frac{1}{M^3})\bigg)M_n^{NS}(Q^2) \nonumber \\
&&+<\Delta p^2>\partial_{p^2}M_n^{NS}(Q^2) \nonumber \\
&&+\frac{2<\bf{p}^2>}{3Q^2}n(n+1)
M_{n+2}^{NS}(Q^2)
\end{eqnarray}
where $M$ is the mass of a nucleon, and for $^{56}Fe$
the parameters of the nuclear model adopted
in Ref.\cite{Kulagin} are
$<\epsilon>\approx -56 ~MeV$,
$<{\bf p}^2>/(2M)\approx 35~MeV$, $<\Delta p^2>_{Fe}\approx -0.17~GeV$
and the derivative $\partial_{p^2}M_n(Q^2)$ takes into account that
the target momentum $p$ can be generally off-mass-shell. This effect
results in the following contribution
\cite{Kulagin}
\begin{equation}
\partial_{p^2}M_n(Q^2)=\partial_{p^2}M_n^{as}+\frac{n}{Q^2}\bigg
(M_n^{NS}+M^2\partial_{p^2}M_n^{as}\bigg)
\end{equation}
which is independent of the nuclear content of the target.
The numerical values of $\partial_{p^2}M_n^{as}$ were also
presented in Ref.\cite{Kulagin}.

Note that the effects of  nuclear corrections in DIS
were also recently studied in Ref.\cite{Spain1} in the case
of $xF_3$  and in Refs.\cite{Spain2},\cite{SL}
in the case of $F_2$  (for  earlier related works, see e.g.,
Ref.\cite{AKV}).
However, in our studies we  concentrate
ourselves on the consideration of the results of Ref.\cite{Kulagin}.

We included the corrections of Eqs.(48)-(49) into our fits and
observed an unacceptable increase of $\chi^2$ value.
We think that this can be related to a
possible asymptotic character of the
$1/M$-expansion in Eq.(48),
since the third term in the
brackets  of the r.h.s. of Eq.(48) becomes
comparable with the first term (which is equal to  unity)
for  $n \sim 8$ used in our fits. Note that
the moments with large $n$ are important in the
reconstruction of the behavior of  $xF_3$  as $x\rightarrow 1$.
This observed feature
necessitates the derivation of the explicit expression
for $M_n^A(Q^2)$ which is not expanded in powers of $1/M$-terms.
It should be added  that the  problem of  possible asymptotic nature
of the power suppressed expansions was mentioned in the case of
Ellis-Jaffe and Bjorken  DIS sum rules  in Ref.\cite{Ioffe}.

Another possibility to explain the non-convergence of our fits
with the nuclear corrections of Eq.(48)   might be related
to the fact
that the parton distribution model for the nuclear SF $xF_3^{^{56}Fe}$
can be different from the canonical model used by us\cite{Kulpr}.
In any case,  we think that  the  study of the problem of
possible influence
of  heavy nuclear effects on the
results  of fits of $xF_3$ data
is  still  on
the agenda.


{\bf Conclusion}

In this work we presented the results of  extractions
of $\alpha_s(M_Z)$  and twist-4 terms from the QCD analysis
of the CCFR data  taking into account definite QCD
corrections at the NNLO and beyond. Within experimental
and theoretical errors our results for
$\alpha_s(M_Z)$ are in agreement with other extractions of this
fundamental parameter, including its world average value
$\alpha_s(M_Z)=0.118 \pm 0.005 $.

Our estimate of the NNLO theoretical uncertainties is based on application
of the [0/2] Pad\'e approach at the N$^3$LO level and the analysis 
of factorization and renormalization scale ambiguities. 
The uncertainties
of our NNLO analysis can be decreased after explicit NNLO calculations of
the NS Altarelli-Parisi kernel.
It should be added, however, that the NLO results, obtained by us
both for the $x$-shape of  twist-4 corrections and for the $\alpha_s(M_Z)$
value are in good agreement with the results of the NLO DGLAP analysis
of the CCFR'97 $xF_3$ and $F_2$ SFs data \cite{AK}, which gives
$\alpha_s(M_Z)=0.1222\pm0.0048~ (exp)\pm 0.0040~(theor)$

As to the twist-4 term, we found that despite the qualitative agreement
of our NLO results with the IRR model prediction, at the NNLO level its
$x$-shape
tends
to decrease and is  stable to
the application of the [0/2] Pad\'e motivated N$^3$LO analysis, 
which is supported by the NNLO fits with fixing factorization 
and renormalization scales as $k_R=k_F=1/4$.

This feature can be related to the fact that
the analysis of the CCFR data cannot distinguish
the twist-4 $1/Q^2$ terms
from the
NNLO perturbative QCD approximations of the Mellin moments.
This possible explanation
is similar to  the conclusions of the LO analysis of the
old less precise neutrino DIS data, made by the authors of
Ref.\cite{Barnett}. It is worth to remind that they
were unable to distinguish between LO logarithmic and $1/Q^2$-behavior
of the QCD contributions to Mellin moments of $xF_3$.
The experimental precision achieved in our days  might move
this effect to  NNLO.
Another possibility is that the inclusion of  NNLO perturbative
QCD contributions  makes the extraction of the $1/Q^2$-corrections
within IRR model approach and by the model-independent way
more problematic (for  discussions of the perturbatively based alternative of
the IRR language within quantum mechanics model see Ref.\cite{PP2}).~

Another related explanation of the decrease of the
twist-4 terms in  NNLO
come from the partial summation of the definite terms of the
asymptotic perturbative QCD series and that the increase of the order
of perturbative QCD analysis effectively suppresses the remaining sum
of the perturbative QCD contribution.
One can hope that  future experiments of the NuTeV
collaboration will allow one to get  new experimental data at the
precision level, necessary for extracting a more detailed information about
higher twist contributions.

{\bf Note added}

After  the technical part of this work was done, we learned
about the work of Ref.\cite{SY} where the NNLO analysis of
$F_2$ SLAC, HERA and BCDMS data was performed both in the singlet and
nonsinglet cases with the help of the method of Bernstein polynomials
\cite{Yn}. The main result of this work is the NNLO value
$\alpha_s(M_Z)=0.1172 \pm 0.0024$, which is in agreement with
our findings. 
In another recent work, the first steps towards
the inclusion of  NNLO corrections to the NS part of $F_2$ 
and modelling the NNLO corrections to the kernel in the $x$-space
were made \cite{VV}. Our analysis of factorization-renormalization 
scale uncertainties confirms their findings.  
It should be also noted, that quite recently the NNLO results 
of Ref.\cite{WZ} were confirmed \cite{MV} with the help 
of other methods. 
We hope that our possible future studies
will allow us to generalize the  Jacobi polynomial NNLO analysis
presented in this work to the case of $F_2$ SF.

{\bf Acknowledgments}

Part of this work was done when one of us (AVS) was visiting
Santiago de Compostela University. He is grateful to his colleagues
for the hospitality in Spain.

We are grateful to A.V.Kotikov for his participation at the first stage
of these  studies and for discussions. The useful comments of
S.I. Alekhin  are gratefully acknowledged.
Special thanks are due to S.A.Kulagin for sharing his
points of view on the current status of  studies of
nuclear effects in DIS.

We would  like to thank G.Altarelli, V.M. Braun, J. Bl\"umlein, K.G. Chetyrkin,
and W. van Neerven for  useful comments on our previous  results, given
in Refs.\cite{KKPS2,KPS}.

We express our special thanks to J. Ch\'yla for careful reading 
the work and valuable advises.

We would also like to acknowledge
S.J. Brodsky, J. Ellis, B.L. Ioffe, L.N. Lipatov, E.A. Kuraev, A.A. Pivovarov
and A.V. Radyushkin  for their interest in the results
of our analysis.

The part of this work was done  when ALK was visiting ICTP, Trieste.
He is grateful to the staff of Abdus Salam International 
Centre for Theoretical Physics  for providing good
working conditions.

This work is done within the scientific
program of the project supported by the Russian Foundation of
Basic  Research,
Grant N 99-01-00091. The work of G.P. was supported by CICYT
(Grant N AEN96-1773) and Xunta de Galicia (Grant N XUGA-20602B98).

\end{document}